\documentclass[11pt,a4paper]{article}

\usepackage{fix-cm}                       % Allow arbitrary font sizes
\usepackage{amsmath, amssymb, amsfonts}  % Standard math packages
\usepackage{graphicx}                     % For figures
\usepackage{hyperref}                     % For links
\usepackage[hypcap=false]{caption}        % Captions for figures/tables
\usepackage{bm}                           % Bold math symbols
  \usepackage[margin=1in]{geometry}  % Page layout

\usepackage[backend=biber,style=numeric, sorting=none]{biblatex}
\usepackage{float}

\numberwithin{equation}{section}

\begin{document}

  % Preprint format (article class)
  \title{Elliptic normal forms and Frobenius--Stickelberger identities for continuous-wave four-wave mixing}
  \author{Graham Hesketh \\
  \small \texttt{gdh1e10@gmail.com}}
  \date{October 25, 2026}
  \maketitle

\begin{abstract}
We study the Hamiltonian and elliptic-function structure of continuous-wave four-wave mixing in nonlinear optical fibres.
For the generic quartic reduction, we express the full complex mode envelopes in terms of Weierstrass $\wp$, $\zeta$, and $\sigma$ functions, retaining pump depletion, cross-phase modulation, and phase mismatch within the coupled-wave model.
The conserved Hamiltonian relation is represented through the Frobenius--Stickelberger determinant identity, connecting the mode products to classical elliptic addition formulas.
For each generic trajectory, a sequence of propagation-dependent gauges, local normalisations, and rescalings, constructed from its conserved quantities and one elliptic quadrature, yields a trajectory of a parameter-free normal form.
The full mode equations retain their four-wave mixing interaction structure under this local normalisation, despite its propagation dependence.
Unlike the generally branched original envelopes, the normal-form solutions are single-valued meromorphic functions of Kronecker theta type.
This formulation exposes mathematical connections with other integrable coupled-mode systems in nonlinear optics.
Numerical comparisons with direct integration of both the complexified system and the physical four-wave mixing equations validate the solution formulas for representative initial data.
\end{abstract}

% ---------- SECTION -------------
% --------------------------------

\section{Introduction}
Four-wave mixing (FWM) is a fundamental parametric process in nonlinear optical fibres, in which the Kerr nonlinearity couples four waves whose frequencies satisfy an energy-conservation relation.
It supports wavelength conversion, parametric amplification, optical signal processing, and quantum light generation \cite{agrawalNonlinearFiberOptics2019,marhicFiberOpticalParametric2007}.
Examples of nonlinear optical processing include FWM-based amplitude regeneration \cite{bottrillOptimisationAmplitudeLimiters2016} and, more recently, the experimental synthesis of reconfigurable optical activation functions for neural-network applications \cite{yangReconfigurableOpticalNonlinear2026}.
These applications motivate the study of FWM and related coupled-mode models; the focus of this paper is their underlying integrable structure.

Analytic solutions for continuous-wave FWM have a substantial history.
Chen and Snyder \cite{chenFourphotonParametricMixing1989} and Chen \cite{chenFourwaveMixingOptical1989} obtained elliptic-function expressions for modal powers, including pump depletion, and Marhic \cite{marhicAnalyticSolutionsPhases2013} gave closed-form phase expressions involving elliptic integrals of the third kind.
Taken together, power and phase formulas already constitute complete analytic descriptions within the assumptions of those models.
The use of Jacobi rather than Weierstrass functions does not change the underlying elliptic character of the problem.
Here we work directly with the complex envelopes, using their logarithmic derivatives to pass from Weierstrass $\wp$ to $\zeta$ and $\sigma$ functions.
This representation makes their mode products and transformations accessible to classical addition identities.

FWM has also been studied from the perspective of Hamiltonian reduction and integrability.
For a particular four-wave mixing Hamiltonian, Odzijewicz and Wawreniuk \cite{odzijewiczIntegrableClassicalQuantum2020} developed classical and quantum reductions, described the classical reduced dynamics on Kummer shapes, and obtained explicit classical and quantum solutions under a frequency-resonance condition.
Their work provides a geometric and algebraic setting for FWM alongside the optical amplitude--phase treatments.
Related background includes exact solutions and phase-space analysis for degenerate FWM \cite{cappelliniThirdorderThreewaveMixing1991}, and reduction and geometric-phase reconstruction for resonant three-wave interactions \cite{alberGeometricPhasesReduction1998}.
Resonant three-wave dynamics also appears in the swinging-spring reduction \cite{holmStepwisePrecessionResonant2002}.
These works concern related models and reconstruction questions; they do not establish the mode-level normalisation studied here.
The present paper examines the complex-envelope and elliptic-identity structure of the fibre model specified below and its complexified Hamiltonian system.

This system belongs to a wider family of continuous-wave coupled-mode equations.
Two-wave and three-wave mixing in quadratic media, polarisation dynamics, and coherent couplers also admit Weierstrass-function representations \cite{heskethGeneralComplexEnvelope2015,heskethCompleteWeierstrassElliptic2026}.
A common polynomial Hamiltonian framework, summarised in \cite[Section~4]{heskethGeneratingIntegrableSystems2026}, places these examples and FWM within a family indexed by mode number, interaction weights, and the polynomial dependence on modal powers.
Their conserved power relations reduce the dynamics to a scalar equation; when its defining polynomial is squarefree, its degree distinguishes the elliptic and higher-genus cases.
The related hyperelliptic construction \cite{zotero-item-855} extends this framework to local sigma-function solutions on smooth odd-degree curves, subject to stated degree and regularity conditions.
Understanding the genus-one examples in detail can therefore inform the less familiar higher-genus systems.
In particular, Weierstrass sigma notation makes visible both the Baker--Akhiezer structure of the genus-one kernels and the determinant identities that have hyperelliptic counterparts.

Our aim is to understand how the conservation laws, complex mode solutions, elliptic addition identities, and changes of variables fit together in FWM.
A particularly striking feature is that a normalisation depending on the propagation variable can preserve the characteristic interaction terms and act on the modal-power exchange equation as though it were only a constant scaling.
The resulting normal form and its meromorphic solutions provide a concrete setting in which to examine these connections.
The principal contribution is thus to the mathematical understanding of an integrable optical system; direct numerical integration remains a practical way to evaluate particular trajectories.

\subsection{Our contribution}
\label{subsec:our-contribution}

We develop two main results for the coupled-wave model considered here.
\begin{enumerate}
\item We derive the full complex mode envelopes in Weierstrass $\wp$, $\zeta$, and $\sigma$ notation for the generic quartic reduction, retaining pump depletion, cross-phase modulation, and phase mismatch within the model.
We then represent the conserved Hamiltonian relation through the Frobenius--Stickelberger determinant, identifying the elliptic addition identity underlying the complete mode products (Sections~\ref{sec:mode-power}--\ref{sec:fs-det-ham-proof}).
The contribution is this explicit envelope and determinant formulation, alongside the established power and phase solutions.

\item Guided by these formulas, we construct a sequence of gauge transformations, local normalisations, and rescalings leading to a parameter-free normal form (Section~\ref{sec:coord-transforms}).
Working locally where the transformations are defined, we show how the interaction structure is retained despite the propagation dependence of the normalisation.
We explain why total interaction degree four is distinguished within this class of square-root normalisations in the original propagation variable (Section~\ref{subsec:four-mode-invariance}).
The normal-form solutions are single-valued meromorphic functions of Kronecker theta type.
\end{enumerate}

Numerical comparisons with direct integration of both the complexified system and the physical fibre equations check the formulas for representative initial data (Section~\ref{sec:num-val}).
Finally, Section~\ref{subsec:kronecker-ba} makes explicit the relation between Kronecker kernels and genus-one Baker--Akhiezer functions and discusses the connection to the higher-genus setting.

% ---------- SECTION -------------
% --------------------------------

\section{The quasi-continuous-wave four-wave mixing system}
In this section we recall the four-wave mixing system from nonlinear fibre optics that forms the focus of our study.
The coupled system of ordinary differential equations in \eqref{eq:fwm-system} is taken from \cite{agrawalNonlinearFiberOptics2019} which references \cite{stolenParametricAmplificationFrequency1982} as the origin of the theory.
The system describes four-wave mixing in the quasi-continuous-wave limit, i.e., when temporal envelope variations are negligible compared to the spatial evolution along the propagation direction, allowing time derivatives to be neglected.
Physically, the equations describe the evolution of four copropagating optical waves that exchange energy through the Kerr nonlinearity.
Each wave experiences self-phase modulation (SPM), where its own intensity induces a phase shift on itself, cross-phase modulation (XPM), where the intensities of other waves induce additional phase shifts, and parametric coupling through the four-wave mixing terms, which enable direct energy transfer between the waves when phase-matching conditions are satisfied:
\begin{align}
\frac{dA_1}{dz} &= \frac{in_2\omega_1}{c}\left[\left(f_{1,1}|A_1|^2 + 2\sum_{k\neq 1} f_{1,k}|A_k|^2\right)A_1 + 2f_{1,2,3,4}A_2^*A_3A_4e^{i\Delta kz}\right], \notag \\
\frac{dA_2}{dz} &= \frac{in_2\omega_2}{c}\left[\left(f_{2,2}|A_2|^2 + 2\sum_{k\neq 2} f_{2,k}|A_k|^2\right)A_2 + 2f_{2,1,3,4}A_1^*A_3A_4e^{i\Delta kz}\right], \notag \\
\frac{dA_3}{dz} &= \frac{in_2\omega_3}{c}\left[\left(f_{3,3}|A_3|^2 + 2\sum_{k\neq 3} f_{3,k}|A_k|^2\right)A_3 + 2f_{3,4,1,2}A_1A_2A_4^*e^{-i\Delta kz}\right], \notag \\
\frac{dA_4}{dz} &= \frac{in_2\omega_4}{c}\left[\left(f_{4,4}|A_4|^2 + 2\sum_{k\neq 4} f_{4,k}|A_k|^2\right)A_4 + 2f_{4,3,1,2}A_1A_2A_3^*e^{-i\Delta kz}\right], \label{eq:fwm-system}
\end{align}
where:
\begin{itemize}
    \item $A_j$ are the slowly-varying complex field amplitudes for waves $j = 1, 2, 3, 4$
    \item $z$ is the propagation distance
    \item $\omega_j$ are the angular frequencies of the respective waves
    \item $c$ is the speed of light in vacuum
    \item $n_2$ is the nonlinear refractive index
    \item $f_{j,j}$ are the self-phase modulation (SPM) coefficients
    \item $f_{j,k}$ (for $j \neq k$) are the cross-phase modulation (XPM) coefficients
    \item $f_{j,k,l,m}$ are the four-wave mixing (FWM) coefficients
    \item $A^*$ denotes complex conjugation
    \item $\Delta k = -\beta_1 - \beta_2 + \beta_3 + \beta_4$ is the phase mismatch
    \item $\beta_j = n(\omega_j)\omega_j/c$ are the propagation constants.
\end{itemize}
The modal overlap integrals are defined in terms of the transverse distribution of the fibre mode $F_j(x,y)$ as:
\begin{align}
\label{eq:overlaps}
f_{j,k,l,m} &= \frac{\langle F_j^* F_k^* F_l F_m \rangle}{\sqrt{\langle |F_j|^2 \rangle \langle |F_k|^2 \rangle \langle |F_l|^2 \rangle \langle |F_m|^2 \rangle}}, \\
f_{j,k} &= f_{k,j} = f_{j,j,k,k}
\end{align}
where $\langle \cdots \rangle = \iint_{-\infty}^{\infty} (\cdots) \, dx\,dy$ denotes the transverse spatial integral. Agrawal says of \eqref{eq:fwm-system} that the equations ``are quite general in the sense that they include the effects of SPM, XPM, and pump depletion on the FWM process'' and stresses that numerical or approximate solutions are commonly utilised.
Herein, they are solved analytically.

% ---------- SECTION -------------
% --------------------------------

\section{The normalised system}
In this section we normalise and generalise the system, transforming it into a form that both simplifies the algebra and broadens the scope of applicability.
From \eqref{eq:overlaps}, we observe the following symmetries among the wave mixing coefficients:

\begin{align}
f_{1,2,3,4} &= |f_{1,2,3,4}|e^{i\varphi}, \notag \\
f_{2,1,3,4} &= f_{1,2,3,4} = |f_{1,2,3,4}|e^{i\varphi}, \notag \\
f_{3,4,1,2} &= f_{1,2,3,4}^* = |f_{1,2,3,4}|e^{-i\varphi}, \notag \\
f_{4,3,1,2} &= f_{1,2,3,4}^* = |f_{1,2,3,4}|e^{-i\varphi}, \label{eq:overlap-symms}
\end{align}
where $\varphi$ is the phase of $f_{1,2,3,4}$.
The symmetries in \eqref{eq:overlap-symms} follow directly from permutation of the first index pair and conjugation in \eqref{eq:overlaps}.
The scalar overlap formula itself is the modelling assumption; vectorial polarisation effects require the corresponding vector overlap model.
Assuming \eqref{eq:overlap-symms}, we can conveniently renormalise the functions, and also absorb the phase $\varphi$ as a global phase rotation of the modes in such a way that the wave mixing coefficients all become equal to one.

To do so, we introduce the following redefinition of physical complex envelopes $A_j, A^*_j$ in terms of abstracted variables $u_j,v_j$.
The complex conjugates $A, A^*$ are distributed among $u,v$ in a particular pattern (note that $u_3, u_4$ involve $A^*_3, A^*_4$ while $v_3, v_4$ involve $A_3, A_4$) so as to later give one product over $u$ and one over $v$ in the wave mixing terms:

\begin{align}
T &= \sqrt{\frac{2\,n_2 \left|f_{1,2,3,4}\right|}{c} \sqrt{ \prod\limits_{k=1}^{4} \omega_k}}, \label{eq:T-def} \\
u_1\left(z\right) &= \frac{T\,\,e^{-i\pi/4}\,e^{-i\varphi/4}}{\sqrt{\omega_{1}}} \,A_{1}\left(z\right)\,e^{i z \beta_{1}}, \notag \\
u_2\left(z\right) &= \frac{T\,e^{-i\pi/4}\,e^{-i\varphi/4}}{\sqrt{\omega_{2}}}\,A_{2}\left(z\right)\,e^{i z \beta_{2}}, \notag \\
u_3\left(z\right) &= \frac{T\,e^{i\pi/4}\,e^{-i\varphi/4}}{\sqrt{\omega_{3}}}\,A^*_{3}\left(z\right)\,e^{-i z \beta_{3}}, \notag \\
u_4\left(z\right) &= \frac{T\,e^{i\pi/4}\,e^{-i\varphi/4}}{\sqrt{\omega_{4}}}\,A^*_{4}\left(z\right)\,e^{-i z \beta_{4}}, \notag \\
v_1\left(z\right) &= \frac{T\,e^{-i\pi/4}\,e^{i\varphi/4}}{\sqrt{\omega_{1}}}\,A^*_{1}\left(z\right)\,e^{-i z \beta_{1}}, \notag \\
v_2\left(z\right) &= \frac{T\,e^{-i\pi/4}\,e^{i\varphi/4}}{\sqrt{\omega_{2}}}\,A^*_{2}\left(z\right)\,e^{-i z \beta_{2}}, \notag \\
v_3\left(z\right) &= \frac{T\,e^{i\pi/4}\,e^{i\varphi/4}}{\sqrt{\omega_{3}}}\,A_{3}\left(z\right)\,e^{i z \beta_{3}}, \notag \\
v_4\left(z\right) &= \frac{T\,e^{i\pi/4}\,e^{i\varphi/4}}{\sqrt{\omega_{4}}}\,A_{4}\left(z\right)\,e^{i z \beta_{4}}. \label{eq:u-v-A}
\end{align}

For physical fibre parameters, $T$ is real and the transformation implies
\begin{equation}
v_j=-is_j u_j^*,\qquad
u_jv_j=-is_j\frac{T^2}{\omega_j}|A_j|^2,
\qquad s_1=s_2=1,\quad s_3=s_4=-1.
\label{eq:physical-real-slice}
\end{equation}
Thus $u_jv_j$ is a scaled, purely imaginary physical power on these solutions; $u_j$ and $v_j$ are not complex conjugates.
We now extend the equations by treating $u_j,v_j$ as independent complex variables.
The physical system is recovered by imposing \eqref{eq:physical-real-slice} on the initial data, a constraint preserved by the equations with physical coefficients.
This complexification provides Hamiltonian conjugate pairs and separates the two interaction products; it does not introduce additional physical degrees of freedom on the constrained subset.
The different formulas for $u_j$ and $v_j$ below reflect these complementary roles and obey the physical conjugation constraint when initialized on that subset.
In all that follows, ``modes'' denotes components of this coupled system, and ``modal power'' denotes the product $u_jv_j$ by analogy; these terms need not refer to transverse fibre modes or real optical powers.

We subsequently introduce parameters $a_j, a_{j,k} \in \mathbb{C}$ which we refer to as propagation constants and phase modulation parameters respectively, again borrowing terminology by analogy.
In the four-wave mixing case these take the values:
\begin{align}
{a}_{j} &= - i\, s_j {\beta}_{j}, \\
{a}_{j,k} &= {a}_{k,j} = - \frac{\left(\delta_{j, k} - 2\right) s_j\, s_k\, {f}_{j,k}\, {\omega}_{j}\, {\omega}_{k}}{2 \left|{{f}_{1,2,3,4}}\right| \sqrt{\prod_{l=1}^{4} {\omega}_{l}}}
\end{align}
with $\delta$ the Kronecker delta function, and $s_j=\pm1$ encodes the signs to match the complex conjugation among $A_j, A^*_j$ with the values given in \eqref{eq:physical-real-slice}.
For fixed initial fields, the propagation constants enter the modal-power evolution through $\sum_j a_j=i\Delta k$; their remaining contributions are carrier phases.
Frequency-independent and frequency-linear shifts of $\beta_j$ cancel from $\Delta k$ under the frequency-conservation relation.
The four-wave mixing system in \eqref{eq:fwm-system} is then a special case of the more general system:
\begin{align}
\frac{d}{d z} u_j{\left(z \right)} &= - \left({a}_{j} + \sum_{k=1}^{4} {a}_{j,k}\,u_k v_k \right) u_j + \prod\limits_{k=1, k \ne j}^{4} v_k, \notag \\
\frac{d}{d z} v_j{\left(z \right)} &= \left({a}_{j} + \sum_{k=1}^{4} {a}_{j,k}\,u_k v_k \right) v_j - \prod\limits_{k=1, k \ne j}^{4} u_k. \label{eq:uv-system}
\end{align}
We take $a_{j,k}=a_{k,j}$ in the complexified system, as required by the Hamiltonian below.
The physical coefficients are a particular instance; complex coefficients also allow related models such as parity-time symmetric systems \cite{heskethGeneralComplexEnvelope2015,sarmaContinuousDiscreteSchrodinger2014}.

The relation to amplitude--phase descriptions can be stated directly.
Where $A_j\ne0$, write $A_j=\sqrt{P_j}\,e^{i\vartheta_j}$; then
\begin{equation}
P_j'=2P_j\operatorname{Re}\!\left(\frac{A_j'}{A_j}\right),\qquad
\vartheta_j'=\operatorname{Im}\!\left(\frac{A_j'}{A_j}\right).
\label{eq:amplitude-phase-correspondence}
\end{equation}
Applying the inverse of \eqref{eq:u-v-A} to our solutions and using these identities recovers the power and phase equations of the same physical model.
With matching coefficients and initial fields, uniqueness of the coupled-wave initial-value problem therefore identifies the resulting fields with those reconstructed from the earlier power and phase solutions.
For example, Marhic's scalar model \cite[Eq.~(1)]{marhicAnalyticSolutionsPhases2013} uses a common nonlinear coefficient $\gamma_{\mathrm{NL}}$ and the scaled distance $Z=\gamma_{\mathrm{NL}}z$.
Comparison with that work requires the same common-coefficient approximation in \eqref{eq:fwm-system}.
Its phase integrals of the third kind and the sigma-function factors used here are alternative representations of the same reconstruction quadratures.
At power zeros the phase chart fails, and the complex fields provide the continuation.

% ---------- SECTION -------------
% --------------------------------

\section{Conserved quantities}

In this section we identify conserved quantities of the system.
The system in \eqref{eq:uv-system} can be formulated as the following canonical Hamiltonian system:
\begin{align}
\label{eq:ham-uv}
H(u_1,\dots, u_4, v_1, \dots, v_4) &= -\sum_{j=1}^{4} a_{j}\, u_j v_j - \frac{1}{2}\,\sum_{j,k=1}^{4} a_{j,k}\, u_j v_j u_k v_k + \prod\limits_{j=1}^{4} u_j + \prod\limits_{j=1}^{4} v_j, \\
\frac{d}{d z} u_j{\left(z \right)} &= \frac{\partial H}{\partial v_j}, \\
\frac{d}{d z} v_j{\left(z \right)} &= -\frac{\partial H}{\partial u_j}, \\
\frac{d}{d z} H &= 0
\end{align}
The conservation of $H$ is to be expected as a consequence of \eqref{eq:ham-uv} lacking an explicit $z$ dependence. The pair $u_j, v_j$ are Hamiltonian conjugates, and each pair represents one of four degrees of freedom in the system.
The modal power $u_j v_j$ evolves according to:
\begin{equation}
\label{eq:duv_j}
\frac{d}{d z} u_j v_j = - \prod\limits_{k=1}^{4} u_k + \prod\limits_{k=1}^{4} v_k
\end{equation}

As the right hand side of \eqref{eq:duv_j} is the same for all $j$,
we may define constants $\gamma_j$ and function $\rho(z)$ such that:
\begin{equation}
\label{eq:uvj-gam-rho}
u_j(z) v_j(z) = \gamma_j - \rho(z)
\end{equation}
from which it follows that all pairwise modal power differences are conserved:
\begin{equation}
\label{eq:power-conserved}
u_j(z) v_j(z) - u_k(z) v_k(z) = \gamma_j - \gamma_k.
\end{equation}
There are six unordered pairwise differences, but only three are independent.
Choosing the fourth mode as a reference gives an independent set
\begin{equation}
C_j=u_jv_j-u_4v_4=\gamma_j-\gamma_4,\qquad j=1,2,3.
\label{eq:independent-power-conserved}
\end{equation}
Every other pairwise difference is a difference of these reference quantities.

With the canonical Poisson bracket
$\{F,G\}=\sum_{l=1}^{4}(F_{u_l}G_{v_l}-F_{v_l}G_{u_l})$,
the modal products $I_j=u_jv_j$ satisfy $\{I_j,I_k\}=0$ and
$\{I_j,H\}=\prod_l v_l-\prod_l u_l$.
Consequently, $H,C_1,C_2,C_3$ are in involution.
They are functionally independent on the generic open set where all $I_j\ne0$ and $\prod_l u_l\ne\prod_l v_l$:
the three reference differences have independent differentials, while the variation
$u_1\mapsto e^t u_1$, $v_1\mapsto e^{-t}v_1$ fixes every $I_j$ and changes $H$ at rate $\prod_l u_l-\prod_l v_l\ne0$.
Thus the complexified four-degree-of-freedom Hamiltonian system is holomorphically Liouville integrable on this open set \cite{arnoldMathematicalMethodsClassical1989}.
The $\gamma_j$ constants can be determined from \eqref{eq:power-conserved} and initial conditions,
however, this provides three equations for four unknowns and thus a choice is available for normalisation.
We choose to impose the constraint that:
\begin{equation}
\sum\limits_{j=1}^{4}\gamma_j = 0,
\end{equation}
after which $\rho(z)$ is minus the mean modal power and $\gamma_j$ is the constant difference between a mode's power and the mean.

% ---------- SECTION -------------
% --------------------------------

\section{Solutions for modal powers in terms of Weierstrass \texorpdfstring{$\wp$}{wp} elliptic functions}\label{sec:mode-power}

In this section we give analytic solutions for the modal powers.
The standing generic assumptions are $d_4\ne0$ and $g_2^3-27g_3^2\ne0$, with nonzero initial envelopes and nonsingular denominators in the displayed reconstruction chart.
Roots and elliptic arguments are selected consistently with the initial data; logarithms and fractional powers are continued on a chosen local branch.
Section~\ref{sec:cubic-case} treats $d_4=0$, $d_3\ne0$ separately; singular curves and boundary charts require limits or separate treatment.
The fundamental Weierstrass elliptic function theory used in this and subsequent sections can be found in \cite{whittakerCourseModernAnalysis2021}.
In order to obtain elliptic function solutions in wave mixing dynamics, the key observation is that the derivative of the modal power in \eqref{eq:duv_j} is proportional to the difference of the two wave mixing product terms, and that the Hamiltonian in \eqref{eq:ham-uv} contains their sum.
We then note the simple but important identity:
\begin{equation}
\label{eq:prod-id}
\left(\prod\limits_{k=1}^{4} u_k - \prod\limits_{k=1}^{4} v_k\right)^2 - \left(\prod\limits_{k=1}^{4} u_k + \prod\limits_{k=1}^{4} v_k\right)^2 = - 4 \prod\limits_{j=1}^{4} u_j v_j
\end{equation}
This enables us to square \eqref{eq:duv_j} and replace wave mixing terms with phase modulation terms, i.e., monomials of $u_j v_j$ which can all be expressed in terms of $\rho(z)$ via \eqref{eq:uvj-gam-rho}.
To proceed in this manner, let us introduce the function $Q$ which represents the phase modulation part of the Hamiltonian and is thus itself expressible as a polynomial of $\rho(z)$:
\begin{align}
Q(u_1(z)v_1(z), \dots, u_4(z)v_4(z)) &= a_0 + \sum_{j=1}^{4} a_{j}\, u_j v_j + \frac{1}{2}\,\sum_{j,k=1}^{4} a_{j,k}\, u_j v_j u_k v_k  \notag  \\
& = \sum_{l=0}^{2}b_l\,\rho\left(z\right)^l \label{eq:Q-uv}
\end{align}
with $a_0=H$ a renaming that indicates the role of $H$ here as the constant term of the polynomial.
We write $\prod_{j=1}^4(\gamma_j-\rho)=\sum_{l=0}^4 c_l\rho^l$, with coefficients listed in Appendix~\ref{app:param-def-appendix}.
Observe that squaring \eqref{eq:duv_j} and substituting \eqref{eq:ham-uv}, \eqref{eq:uvj-gam-rho}, \eqref{eq:prod-id}, and \eqref{eq:Q-uv} gives:
\begin{align}
\left(\frac{d}{d z} \rho{\left(z \right)}\right)^{2} &= Q^{2}{\left({\gamma}_{1} - \rho{\left(z \right)}, \dots, {\gamma}_{4}  - \rho{\left(z \right)}\right)} - 4 \prod_{j=1}^{4} \left({\gamma}_{j} - \rho{\left(z \right)}\right), \label{eq:drho-sqrd-1} \\
\left(\frac{d}{d z} \rho{\left(z \right)}\right)^{2} &=  \left(\sum_{l=0}^{2}b_l\,\rho\left(z\right)^l\right)^{2} - 4 \prod_{j=1}^{4} \left({\gamma}_{j} - \rho{\left(z \right)}\right), \label{eq:drho-sqrd-2} \\
\left(\frac{d}{d z} \rho{\left(z \right)}\right)^{2} &= \sum_{l=0}^{4}d_l\,\rho\left(z\right)^l, \label{eq:drho-sqrd-3} \\
\left(\frac{d}{d z} \rho{\left(z \right)}\right)^{2} &= d_4 \prod_{l=1}^{4} \left(\rho\left(z\right) -  \lambda_l\right), \label{eq:drho-sqrd-4}
\end{align}
where $b_l$ and $d_l$ are given in terms of other parameters and initial conditions in Appendices \ref{app:param-def-appendix} and \ref{app:init-conds}, and where $\lambda_l$ are the roots of the quartic polynomial in $\rho(z)$, i.e. $0=\sum_{k=0}^{4}d_k\,\lambda_{l}^k$.
Equation \eqref{eq:drho-sqrd-4} follows algebraically from the conserved quantities.
Solving this equation, and fixing the elliptic arguments from initial data, still involves inversion of an elliptic integral; the Weierstrass representation does not remove that step.
To solve it, we now transform \eqref{eq:drho-sqrd-4} from quartic to the standard cubic form of the Weierstrass $\wp$ function using the classical trick which can be conceptualised in three steps \cite{whittakerCourseModernAnalysis2021}:
\begin{equation}
\label{eq:wp-4-to-3}
\begin{array}{l rcl}
\text{(1) shift by any root so RHS is 0 when $q(z)=0$}
& \rho(z) &=& q(z) + \lambda_1,\\
\text{(2) invert to reduce the quartic to a cubic}
& s(z) &=& \dfrac{1}{q(z)},\\
\text{(3) shift and scale to match Weierstrass coefficients}
& w(z) &=& C_1 s(z) + C_0.
\end{array}
\end{equation}

In the generic quartic case we assume four distinct roots and choose one, denoted $\lambda_1$.
The procedure sketched in \eqref{eq:wp-4-to-3} is then implemented in the following single transformation:
\begin{align}
\rho{\left(z \right)} &= {\lambda}_{1} + \frac{d_4}{- 4 w{\left(z \right)} \prod_{l=1}^{3} {\Omega}_{l}  + \frac{d_4}{3} \sum_{l=1}^{3} {\Omega}_{l} }, \label{eq:rho-wp-1} \\
\Omega_l &= \frac{1}{\lambda_{l+1} - \lambda_1}, \notag \\
\left(\frac{d}{d z} w{\left(z \right)}\right)^{2} &= 4\,w{\left(z \right)}^3 - g_2\,w{\left(z \right)} - g_3, \label{eq:dwp} \\
g_{2} &= {d}_{0} {d}_{4} - \frac{{d}_{1} {d}_{3}}{4} + \frac{{d}_{2}^{2}}{12}, \label{eq:g2-d} \\
g_{3} &= \frac{{d}_{0} {d}_{2} {d}_{4}}{6} - \frac{{d}_{0} {d}_{3}^{2}}{16} - \frac{{d}_{1}^{2} {d}_{4}}{16} + \frac{{d}_{1} {d}_{2} {d}_{3}}{48} - \frac{{d}_{2}^{3}}{216} \label{eq:3-d}
\end{align}
where the constants $g_2$ and $g_3$ are known as Weierstrass elliptic invariants.
Equation \eqref{eq:dwp} defines the Weierstrass elliptic $\wp$ function and the solution is:
\begin{equation}
w(z) = \wp (z - z_0, g_2, g_3)
\end{equation}
where $z_0$ is a constant chosen to match initial conditions.
This constant $z_0$, and others that we will now introduce, can be obtained by inverting $\wp$ using an elliptic integral \cite{whittakerCourseModernAnalysis2021}.
As $\wp$ is an even function, it is necessary to also specify a corresponding condition for the derivative when inverting to fix the sign, i.e., to find $z$ from known $x,y$ we give conditions such as $\wp\left(z\right)=x, \wp'\left(z\right)=y$, where $\wp'$ is the derivative of $\wp$ known as Weierstrass p-prime.
From elliptic function theory, the points obtained during such an inversion are determined modulo the period lattice of the doubly periodic $\wp$, which is a two-dimensional generalisation of saying they are only uniquely determined within a period of oscillation.
In addition to $z_0$, we introduce two other types of special points that characterize the solution structure and enable the use of elliptic function identities to solve for the modes in Section \ref{sec:mode-sols-u-v}.
The two poles of $\rho(z)$ modulo the lattice are $z_0\pm z_1$, where the denominator in \eqref{eq:rho-wp-1} vanishes.
We select $\mu_j$ as the zero of $v_j$, so $\rho(\mu_j)=\gamma_j$ and $\rho'(\mu_j)=Q(\gamma_j)$; the corresponding zero of $u_j$ is $2z_0-\mu_j$ modulo the lattice.
The sign of $\sqrt{d_4}$ in the derivative condition for $z_1$ fixes a convention; reversing it exchanges $z_1$ and $-z_1$.
These special points are defined implicitly through their Weierstrass function values:
\begin{align}
\wp{\left(z_{0}\right)} &= \frac{{d}_{2}}{12} + \frac{{d}_{3} {\lambda}_{1}}{4} + \frac{{d}_{4} {\lambda}_{1}^{2}}{2} + \frac{- {d}_{1} - 2 {d}_{2} {\lambda}_{1} - 3 {d}_{3} {\lambda}_{1}^{2} - 4 {d}_{4} {\lambda}_{1}^{3}}{4 \left(- \rho{\left(0 \right)} + {\lambda}_{1}\right)}, \notag \\
\wp'{\left(z_{0}\right)} &= \frac{\left({d}_{1} + 2 {d}_{2} {\lambda}_{1} + 3 {d}_{3} {\lambda}_{1}^{2} + 4 {d}_{4} {\lambda}_{1}^{3}\right) }{4 \left(\rho{\left(0 \right)} - {\lambda}_{1}\right)^{2}}\left. \frac{d}{d z} \rho{\left(z \right)} \right|_{\substack{ z=0 }}, \notag \\
\wp{\left(z_{1}\right)} &= \frac{{d}_{2}}{12} + \frac{{d}_{3} {\lambda}_{1}}{4} + \frac{{d}_{4} {\lambda}_{1}^{2}}{2}, \notag \\
\wp'{\left(z_{1}\right)} &= \frac{\left(- {d}_{1} - 2 {d}_{2} {\lambda}_{1} - 3 {d}_{3} {\lambda}_{1}^{2} - 4 {d}_{4} {\lambda}_{1}^{3}\right) \sqrt{{d}_{4}}}{4}, \notag \\
\wp{\left({\mu}_{j} - z_{0}\right)} &= \frac{{d}_{2}}{12} + \frac{{d}_{3} {\lambda}_{1}}{4} + \frac{{d}_{4} {\lambda}_{1}^{2}}{2} - \frac{- {d}_{1} - 2 {d}_{2} {\lambda}_{1} - 3 {d}_{3} {\lambda}_{1}^{2} - 4 {d}_{4} {\lambda}_{1}^{3}}{4 \left({\gamma}_{j} - {\lambda}_{1}\right)}, \notag \\
\wp'{\left({\mu}_{j}- z_{0}\right)} &= - \frac{\left({b}_{0} + {b}_{1} {\gamma}_{j} + {b}_{2} {\gamma}_{j}^{2}\right) \left({d}_{1} + 2 {d}_{2} {\lambda}_{1} + 3 {d}_{3} {\lambda}_{1}^{2} + 4 {d}_{4} {\lambda}_{1}^{3}\right)}{4 \left({\gamma}_{j} - {\lambda}_{1}\right)^{2}}. \label{eq:wp-points}
\end{align}

The solutions for modal powers are then found from \eqref{eq:uvj-gam-rho} and \eqref{eq:rho-wp-1} to be:
\begin{align}
u_j{\left(z\right)} v_j{\left(z\right)} &= \rho(\mu_j) - \rho(z) \notag \\
&= \frac{\wp'{\left(z_{1}\right)}}{ \sqrt{{d}_{4}} \left(\wp{\left(z_{1}\right)} - \wp{\left({\mu}_{j} - z_{0}\right)}\right)} - \frac{\wp'{\left(z_{1}\right)}}{\sqrt{{d}_{4}} \left(\wp{\left(z_{1}\right)} - \wp{\left(z - z_{0}\right)}\right) } \notag \\
&= \frac{\wp'{\left(z_{1}\right)} }{ \sqrt{{d}_{4}} \left(\wp{\left(z_{1}\right)} - \wp{\left({\mu}_{j}- z_{0}\right)}\right)} \frac{ \left(\wp{\left(z - z_{0}\right)} - \wp{\left({\mu}_{j} - z_{0}\right)}\right) }{ \left(\wp{\left(z - z_{0}\right)} - \wp{\left(z_{1}\right)}\right) }. \label{eq:uv-wp}
\end{align}

% ---------- SECTION -------------
% --------------------------------

\section{Solutions for modes in terms of Weierstrass \texorpdfstring{$\sigma$}{sigma}, \texorpdfstring{$\zeta$}{zeta} functions} \label{sec:mode-sols-u-v}

Having obtained the modal power solutions in the previous section, we now proceed to find solutions for the individual modes $u_j(z)$ and $v_j(z)$ themselves.
Direct substitution of \eqref{eq:ham-uv}, \eqref{eq:duv_j}, \eqref{eq:uvj-gam-rho}, and \eqref{eq:power-conserved} into \eqref{eq:uv-system} shows that the equations can be recast in terms of logarithmic derivatives:
\begin{align}
\frac{\frac{\mathrm{d}}{\mathrm{d}z} u_j{\left(z\right)}}{u_j{\left(z\right)}} &= \frac{1}{2}\frac{\rho'{\left(z \right)} - \rho'{\left({\mu}_{j} \right)}}{\rho{\left(z \right)} - \rho{\left({\mu}_{j} \right)}} + \rho{\left(z \right)} {\Lambda}_{1,j} + {\Lambda}_{0,j}, \notag \\
\frac{\frac{\mathrm{d}}{\mathrm{d}z} v_j{\left(z\right)}}{v_j{\left(z\right)}} &= \frac{1}{2}\frac{\rho'{\left(z \right)} + \rho'{\left({\mu}_{j} \right)}}{\rho{\left(z \right)} - \rho{\left({\mu}_{j} \right)}} - \rho{\left(z \right)} {\Lambda}_{1,j} - {\Lambda}_{0,j}, \label{eq:dlog-u-v-rho} \\
{\Lambda}_{0,j} &= - {a}_{j} - \frac{{\gamma}_{j}}{4} \sum\limits_{k,l=1}^{4} {a}_{k,l} - \sum_{k=1}^{4} {a}_{j,k} {\gamma}_{k} + \frac{1}{2}\sum_{k=1}^{4} {\gamma}_{k} \sum_{l=1}^{4} {a}_{k,l} + \frac{1}{2}\sum_{k=1}^{4} {a}_{k}, \\
{\Lambda}_{1,j} &= \sum_{k=1}^{4} {a}_{j,k} - \frac{1}{4}\sum\limits_{k,l=1}^{4}  {a}_{k,l},
\end{align}
where the right-hand side is expressed purely in terms of $\rho$ and its derivative $\rho'$.
This logarithmic derivative form is advantageous because it can be integrated directly to yield solutions in terms of Weierstrass $\sigma$ functions, as we now show.

We substitute the modal power solution \eqref{eq:uv-wp} into \eqref{eq:dlog-u-v-rho} and apply the elliptic function identity:
\begin{equation}
\frac{\wp'{\left(x,g_{2},g_{3} \right)}}{\wp{\left(x,g_{2},g_{3} \right)} - \wp{\left(y,g_{2},g_{3} \right)}} = \zeta{\left(x + y,g_{2},g_{3} \right)} + \zeta{\left(x - y,g_{2},g_{3} \right)} - 2 \zeta{\left(x,g_{2},g_{3} \right)}
\end{equation}
which allows us to express the right-hand side in terms of Weierstrass $\zeta$ functions and constants:
\begin{align}
\frac{\frac{\mathrm{d}}{\mathrm{d}z} u_j{\left(z\right)}}{u_j{\left(z\right)}} =& \frac{\left(\zeta{\left(z - z_{0} + z_{1} \right)} - 2 \zeta{\left(z_{1} \right)} - \zeta{\left(z - z_{0} - z_{1} \right)}\right) {\Lambda}_{1,j}}{\sqrt{{d}_{4}}} \nonumber \\[6pt]
& + \zeta{\left(z - 2 z_{0} + {\mu}_{j} \right)} - \frac{\zeta{\left(z - z_{0} - z_{1} \right)}}{2} - \frac{\zeta{\left(z - z_{0} + z_{1} \right)}}{2} \nonumber \\[6pt]
& - \frac{\zeta{\left({\mu}_{j} - z_{0} - z_{1}\right)}}{2} - \frac{\zeta{\left({\mu}_{j} - z_{0} + z_{1}\right)}}{2} + {\Lambda}_{0,j} + {\Lambda}_{1,j} {\lambda}_{1} \notag \\[12pt]
\frac{\frac{\mathrm{d}}{\mathrm{d}z} v_j{\left(z\right)}}{v_j{\left(z\right)}} =& - \frac{\left(\zeta{\left(z - z_{0} + z_{1} \right)} - 2 \zeta{\left(z_{1} \right)} - \zeta{\left(z - z_{0} - z_{1} \right)}\right) {\Lambda}_{1,j}}{\sqrt{{d}_{4}}} \nonumber \\[6pt]
& + \zeta{\left(z - {\mu}_{j} \right)} - \frac{\zeta{\left(z - z_{0} - z_{1} \right)}}{2} - \frac{\zeta{\left(z - z_{0} + z_{1} \right)}}{2} \nonumber \\[6pt]
& + \frac{\zeta{\left({\mu}_{j} - z_{0} - z_{1}\right)}}{2} + \frac{\zeta{\left({\mu}_{j} - z_{0} + z_{1} \right)}}{2} - {\Lambda}_{0,j} - {\Lambda}_{1,j} {\lambda}_{1}. \label{eq:dlog-u-v-zeta}
\end{align}
Equations \eqref{eq:dlog-u-v-zeta} can be integrated by noting that the Weierstrass $\zeta$ function is the logarithmic derivative of the Weierstrass $\sigma$ function:
\begin{equation}
\label{eq:zeta-dlog_sigma}
\zeta\left(z, g_2, g_3\right)=\frac{\mathrm{d}}{\mathrm{d}z} \log{\left(\sigma\left(z, g_2, g_3\right)\right)}.
\end{equation}
Performing the integration and taking exponentials gives the first main result of this paper, the complete analytic solutions for the complex mode envelopes $u_j, v_j$ without decomposition into amplitude and phase:
\begin{align}
u_j{\left(z\right)} &= \frac{\alpha_j\,\sqrt{W_j} \,\sigma{\left(z - 2 z_{0} + {\mu}_{j} \right)} \exp\left(z {r}_{0,j} + \log{\left(\frac{\sigma{\left(z - z_{0} + z_{1} \right)}}{\sigma{\left(z - z_{0} - z_{1} \right)}} \right)} {r}_{1,j}\right)}{\sqrt{\wp{\left(z_{1} \right)} - \wp{\left(z - z_{0} \right)}} \sigma{\left({\mu}_{j} - z_{0}\right)} \sigma{\left(z - z_{0} \right)}}, \notag \\[12pt]
v_j{\left(z\right)} &= \frac{\sqrt{W_j} \, \sigma{\left(z - {\mu}_{j} \right)} \exp\left(- z {r}_{0,j} - \log{\left(\frac{\sigma{\left(z - z_{0} + z_{1} \right)}}{\sigma{\left(z - z_{0} - z_{1} \right)}} \right)} {r}_{1,j} \right)}{\alpha_j\,\sqrt{\wp{\left(z_{1} \right)} - \wp{\left(z - z_{0} \right)}} \sigma{\left({\mu}_{j} - z_{0}\right)} \sigma{\left(z - z_{0} \right)}} \label{eq:u-v-quartic}
\end{align}
where the branch of the logarithm is chosen continuously along the integration path from $z=0$, $\alpha_j$ is the integration constant that can be fixed by initial conditions to capture any phase offset between a mode and its conjugate, and the other constants are:
\begin{align}
W_j &= \gamma_j-\lambda_1, \\[6pt]
r_{0,j} &= {\Lambda}_{0,j} + {\Lambda}_{1,j} {\lambda}_{1} - \frac{2 \zeta{\left(z_{1} \right)} {\Lambda}_{1,j}}{\sqrt{{d}_{4}}} - \frac{\zeta{\left({\mu}_{j} - z_{0} - z_{1}\right)}}{2} - \frac{\zeta{\left({\mu}_{j} - z_{0} + z_{1}\right)}}{2}, \\[6pt]
r_{1,j} &= \frac{{\Lambda}_{1,j}}{\sqrt{{d}_{4}}}.
\end{align}
These solutions apply in the generic quartic regime of the stated coupled-wave model, retaining pump depletion, cross-phase modulation, and phase mismatch.
The reconstruction is understood on the generic branch chart specified in Section~\ref{sec:mode-power}.
The four $\alpha_j$ are constrained: their product is fixed by $2\prod_j u_j(0)=\rho'(0)+Q(\rho(0))$.
Thus the continuous constants comprise three independent $\gamma_j$, the Hamiltonian, $z_0$, and three independent envelope constants, giving eight complex parameters, as required.
The definitions also give $\sum_j\Lambda_{1,j}=\sum_jr_{1,j}=0$, so the logarithmic factors cancel from the complete mode products.

% ---------- SECTION -------------
% --------------------------------

\section{The Frobenius--Stickelberger determinant}
\label{sec:fs-det-ham-proof}

The Frobenius--Stickelberger (FS) determinant formula is an elliptic function identity relating products of $\sigma$ functions to determinants of derivatives of $\wp$ \cite{whittakerCourseModernAnalysis2021, frobeniusZurTheorieElliptischen1877}.
In this section we relate the conserved Hamiltonian equation to this classical identity in the elliptic solution representation.

Our starting point is equation \eqref{eq:fs-det-start}, which relates the wave mixing product to the phase modulation terms via the Hamiltonian.
We will show that this equation can be transformed into the FS determinant formula by treating the left-hand side (the product of modes) and right-hand side (the polynomial in $\rho$) separately, each ultimately expressed in terms of Weierstrass functions.

We are interested in the $4\times4$ case of FS in the form:
\begin{align}
\label{eq:fs-det-m}
\frac{C\,\sigma{\left(z + {\nu}_{1}\right)} \sigma{\left(z + {\nu}_{2}\right)} \sigma{\left(z + {\nu}_{3}\right)} \sigma{\left(z - {\nu}_{1} - {\nu}_{2} - {\nu}_{3}\right)}}{\sigma^{4}{\left(z\right)}} & = \det M
\end{align}
where:
\begin{align}
C = &\frac{12  \sigma{\left({\nu}_{1} - {\nu}_{2}\right)} \sigma{\left({\nu}_{1} - {\nu}_{3}\right)} \sigma{\left({\nu}_{2} - {\nu}_{3}\right)} }{\sigma^{4}{\left({\nu}_{1}\right)} \sigma^{4}{\left({\nu}_{2}\right)} \sigma^{4}{\left({\nu}_{3}\right)}} \\[6pt]
\det M = &
\begin{vmatrix}
1 & \wp(z) & \frac{\partial \wp(z)}{\partial z} & \frac{\partial^2 \wp(z)}{\partial z^2} \\
1 & \wp(\nu_1) & -\frac{\partial \wp(\nu_1)}{\partial \nu_1} & \frac{\partial^2 \wp(\nu_1)}{\partial \nu_1^2} \\
1 & \wp(\nu_2) & -\frac{\partial \wp(\nu_2)}{\partial \nu_2} & \frac{\partial^2 \wp(\nu_2)}{\partial \nu_2^2} \\
1 & \wp(\nu_3) & -\frac{\partial \wp(\nu_3)}{\partial \nu_3} & \frac{\partial^2 \wp(\nu_3)}{\partial \nu_3^2} \\
\end{vmatrix}  \notag \\[6pt]
= & - 6 \left(\wp{\left({\nu}_{1}\right)} - \wp{\left(z\right)}\right) \left(\wp{\left({\nu}_{2}\right)} - \wp{\left(z\right)}\right) \left(\wp{\left({\nu}_{1}\right)} - \wp{\left({\nu}_{2}\right)}\right) \wp'{\left({\nu}_{3}\right)} \notag \\
& + 6 \left(\wp{\left({\nu}_{1}\right)} - \wp{\left(z\right)}\right) \left(\wp{\left({\nu}_{3}\right)} - \wp{\left(z\right)}\right) \left(\wp{\left({\nu}_{1}\right)} - \wp{\left({\nu}_{3}\right)}\right) \wp'{\left({\nu}_{2}\right)}  \notag \\
& - 6 \left(\wp{\left({\nu}_{2}\right)} - \wp{\left(z\right)}\right) \left(\wp{\left({\nu}_{3}\right)} - \wp{\left(z\right)}\right) \left(\wp{\left({\nu}_{2}\right)} - \wp{\left({\nu}_{3}\right)}\right) \wp'{\left({\nu}_{1}\right)}  \notag \\
& - 6 \left(\wp{\left({\nu}_{1}\right)} - \wp{\left({\nu}_{2}\right)}\right) \left(\wp{\left({\nu}_{1}\right)} - \wp{\left({\nu}_{3}\right)}\right) \left(\wp{\left({\nu}_{2}\right)} - \wp{\left({\nu}_{3}\right)}\right) \wp'{\left(z\right)}
\end{align}

The starting point for us is the following relation established via \eqref{eq:ham-uv}, \eqref{eq:duv_j}, and \eqref{eq:Q-uv}, in which any $j$ applies in the RHS modal power derivative:
\begin{align}
2 \prod_{k=1}^{4} u_k &= {a}_{0} + \sum_{k=1}^{4} {a}_{k} u_k v_k  + \frac{1}{2}\sum\limits_{k,l=1}^{4} {a}_{k,l} u_k v_k u_l v_l - \frac{d}{d z}\left(u_j v_j\right), \notag \\
&=\rho' + \sum_{l=0}^{2}b_l\,\rho\left(z\right)^l. \label{eq:fs-det-start}
\end{align}
We will show how \eqref{eq:fs-det-start} can be transformed into \eqref{eq:fs-det-m}.
Geometrically, $\prod_j u_j$ is elliptic in $x=z-z_0$, with four zeros at $-\nu_j$ and double poles at $x=\pm z_1$ in the generic noncoincident case.
Abel's divisor condition therefore gives $\sum_j\nu_j\equiv0$ modulo the lattice.
The explicit quasiperiod calculation below also determines the exponential normalisation.

\subsection{The left hand side}
From \eqref{eq:uv-wp}, the right hand side of \eqref{eq:fs-det-start} is a doubly periodic elliptic function,
so too then is the left hand side.
Choose primitive generators $2\omega_1,2\omega_3$ of the lattice of $\wp(\cdot;g_2,g_3)$, ordered so that $\operatorname{Im}(\omega_3/\omega_1)>0$, and put $\eta_j=\zeta(\omega_j)$ for $j=1,3$.
These half-periods are unrelated to the optical angular frequencies denoted by $\omega_j$ in Section~2.
For integers $m,n$ we then have:
\begin{align}
\prod_{j=1}^{4} \frac{u_j{\left(z\right)}}{u_j{\left(2 m {\omega}_{3} + 2 n {\omega}_{1} + z\right)}} & = 1. \label{eq:u-prod-is-one}
\end{align}
We substitute our solution \eqref{eq:u-v-quartic} into \eqref{eq:u-prod-is-one} and use the quasi-periodicity of $\sigma$:
\begin{align}
\sigma{\left(2 m {\omega}_{3} + 2 n {\omega}_{1} + z\right)} = \left(-1\right)^{m n + m + n} \sigma{\left(z\right)} e^{\left(2 m {\omega}_{3} + 2 n {\omega}_{1} + 2 z\right) \left(m\eta_3+n\eta_1\right)} \label{eq:quasi-sigma}
\end{align}
to reduce the product in \eqref{eq:u-prod-is-one} to an exponential which, as it is equal to one, must have argument $2\pi i \mathrm{N}(n,m)$, with $\mathrm{N}(n,m)$ an integer that varies with $n,m$.
This leads to:
\begin{align}
2 i \pi \mathrm{N}{\left(n,m \right)} &= - \left(2 m {\omega}_{3} + 2 n {\omega}_{1}\right) \sum_{j=1}^{4} {r}_{0,j} - 2 \left(m\eta_3+n\eta_1\right) \sum_{j=1}^{4} {\nu}_{j}, \label{eq:exp-arg-is-one} \\
\sum_{j=1}^{4} {\nu}_{j} &= 2 \mathrm{N}{\left(0,1 \right)} {\omega}_{1} - 2 \mathrm{N}{\left(1,0 \right)} {\omega}_{3} = 0 \pmod{\text{lattice}}, \label{eq:nuj-sum-zero}  \\
\sum_{j=1}^{4} {r}_{0,j} &= 2\mathrm{N}(1,0)\eta_3-2\mathrm{N}(0,1)\eta_1 \label{eq:roj-quasiperiod}
\end{align}
where $\nu_j=\mu_j-z_0$, and in moving from \eqref{eq:exp-arg-is-one} to \eqref{eq:nuj-sum-zero} and \eqref{eq:roj-quasiperiod} we made use of the identity $\eta_1\omega_3-\eta_3\omega_1=i\pi/2$.
Equivalently, for integers $p=\mathrm{N}(0,1)$ and $q=-\mathrm{N}(1,0)$, the two sums are $\sum_j\nu_j=2p\omega_1+2q\omega_3$ and $\sum_jr_{0,j}=-2p\eta_1-2q\eta_3$. Thus equations \eqref{eq:quasi-sigma}, \eqref{eq:nuj-sum-zero} and \eqref{eq:roj-quasiperiod} allow us to make the following substitution in the left hand side of \eqref{eq:fs-det-start}:
\begin{align}
\frac{\sigma{\left(z + {\nu}_{4} \right)}}{\sigma{\left({\nu}_{4} \right)}}\,\exp{\left(z \sum\limits_{j=1}^{4} {r}_{0,j}\right)} = - \frac{\sigma{\left(z - {\nu}_{1} - {\nu}_{2} - {\nu}_{3} \right)}}{\sigma{\left({\nu}_{1} + {\nu}_{2} + {\nu}_{3} \right)}}.
\end{align}

\subsection{The right hand side}
For the quadratic polynomial $Q(\rho)=\sum_{l=0}^{2}b_l\rho^l$ and distinct nodes $\gamma_1,\gamma_2,\gamma_3$, Waring-Lagrange interpolation gives:
\begin{align}
\sum_{l=0}^{2}b_l\,\rho\left(z\right)^l =\, &\frac{\left({\gamma}_{1} - \rho{\left(z \right)}\right) \left({\gamma}_{2} - \rho{\left(z \right)}\right)}{\left({\gamma}_{1} - {\gamma}_{3}\right) \left({\gamma}_{2} - {\gamma}_{3}\right)} \sum\limits_{l=0}^{2}b_l\,\gamma_3^l + \notag \\
&\frac{\left({\gamma}_{1} - \rho{\left(z \right)}\right) \left({\gamma}_{3} - \rho{\left(z \right)}\right)}{\left({\gamma}_{1} - {\gamma}_{2}\right) \left({\gamma}_{3} - {\gamma}_{2}\right)} \sum\limits_{l=0}^{2}b_l\,\gamma_2^l + \notag \\
&\frac{\left({\gamma}_{2} - \rho{\left(z \right)}\right) \left({\gamma}_{3} - \rho{\left(z \right)}\right)}{\left({\gamma}_{1} - {\gamma}_{2}\right) \left({\gamma}_{1} - {\gamma}_{3}\right)} \sum\limits_{l=0}^{2}b_l\,\gamma_1^l \label{eq:waring-lagrange}
\end{align}
Here $\gamma_j=\rho(\mu_j)$, and the derivative values prescribed in \eqref{eq:wp-points} select the sheets so that
\begin{align}
\sum\limits_{l=0}^{2}b_l\,\gamma_j^l = \rho'\left(\mu_j\right).
\end{align}
This sign is fixed by the sheet choice, rather than by the squared equation \eqref{eq:drho-sqrd-2} alone.
We can then use \eqref{eq:uv-wp} to express $\rho$ in terms of $\wp$ so that we can substitute all variables in \eqref{eq:waring-lagrange} for values in terms of $\wp, \wp'$.
We now substitute $z\mapsto z+z_0$ and write $\nu_j=\mu_j-z_0$, so that the original elliptic argument $z-z_0$ becomes $z$.
We divide both sides of the translated relation \eqref{eq:fs-det-start} by:
\begin{align}
\frac{\wp'{\left(z_{1}\right)}}{\sqrt{{d}_{4}} \left(\wp{\left(z_{1}\right)} - \wp{\left(z\right)}\right)^2 }
\end{align}
to obtain a polynomial expression in $\wp(z)$ and $\wp'(z)$.
For clarity, put $w=\wp(z)$, $w_j=\wp(\nu_j)$, $y_j=\wp'(\nu_j)$,
$A=\wp(z_1)$, and $K=\mathcal F'(\lambda_1)/4$, where $\mathcal F(\rho)=\sum_{l=0}^4d_l\rho^l$.
The quartic reduction gives $\rho=\lambda_1+K/(w-A)$.
The divisor displayed above is $B=-K/(w-A)^2$, so $\rho'/B=\wp'(z)$ and $Q(\rho)/B$ is quadratic in $w$.
By the selected sheets it equals $y_j$ at $w=w_j$; hence
\begin{equation}
\frac{\rho'+Q(\rho)}{B}=\wp'(z)+L(w),\qquad
L(w)=\sum_{j=1}^{3}y_j
\prod_{\substack{k=1\\k\ne j}}^{3}\frac{w-w_k}{w_j-w_k}.
\end{equation}
Expanding the determinant in \eqref{eq:fs-det-m} gives exactly
\begin{equation}
\det M=-6\Delta\,[\wp'(z)+L(w)],\qquad
\Delta=(w_1-w_2)(w_1-w_3)(w_2-w_3).
\end{equation}
The left-hand side, after the exponential absorption above and division by $B$, is a constant multiple of the sigma ratio in \eqref{eq:fs-det-m}.
Multiplying by $\sigma(z)^4$ and taking $z\to0$ fixes this constant and gives the displayed coefficient $C$.
This calculation is made for distinct interpolation nodes and nonzero denominators; coincident cases require limits where these exist.

Thus the conserved Hamiltonian product relation is represented by the four-point Frobenius--Stickelberger identity in these elliptic solution coordinates.
The derivation uses Hamiltonian conservation; it does not derive that conservation independently from the determinant formula.
The first three sheet values determine the quadratic interpolant $L$; the fourth satisfies $\wp'(\nu_4)=L(\wp(\nu_4))$.
For distinct points this is the fourth intersection condition for the elliptic curve and that quadratic, equivalently the four-point divisor condition.
It expresses the compatibility of the four selected sheets with one conserved Hamiltonian polynomial.
The FS determinant framework may provide tools for generalising these results to higher-order wave mixing processes or systems with additional degrees of freedom.

% ---------- SECTION -------------
% --------------------------------

\section{When quartic terms cancel through special parameter values}
\label{sec:cubic-case}

The condition $d_4=b_2^2-4=0$ is a constraint on the original system coefficients, independent of its initial data.
For common overlaps and nearby optical frequencies, the physical model instead has $b_2\simeq-1$ and $d_4\simeq-3$; the cubic case is a distinct coefficient regime.
For $d_3\ne0$ and a nonsingular cubic, put $\upsilon_j=\mu_j-z_0+z_1$, where $z_1$ is a nonzero half-period.
Then
\begin{equation}
\rho(z)=\frac{4}{d_3}\wp(z-z_0+z_1)-\frac{d_2}{3d_3}.
\end{equation}
Integrating \eqref{eq:dlog-u-v-rho} gives
\begin{align}
u_j(z)&=\frac{2\alpha_j\sigma(z-2z_0+2z_1+\mu_j)}{\sqrt{d_3}\,\sigma(\upsilon_j)\sigma(z-z_0+z_1)}e^{E_j(z)},\notag\\
v_j(z)&=\frac{2\sigma(z-\mu_j)}{\alpha_j\sqrt{d_3}\,\sigma(\upsilon_j)\sigma(z-z_0+z_1)}e^{-E_j(z)},\notag\\
E_j(z)&=z\left(\Lambda_{0,j}-\frac{d_2\Lambda_{1,j}}{3d_3}-\zeta(\upsilon_j)\right)
-\frac{4\Lambda_{1,j}}{d_3}\zeta(z-z_0+z_1).
\label{eq:u-v-cubic}
\end{align}
The constants $\alpha_j$ are fixed from the initial envelopes using this formula.
The extra zeta exponential is required because integrating $\Lambda_{1,j}\rho$ contributes $-4\Lambda_{1,j}\zeta(z-z_0+z_1)/d_3$.
The modal powers are
\begin{equation}
u_j(z)v_j(z)=\frac{4}{d_3}\bigl[\wp(\upsilon_j)-\wp(z-z_0+z_1)\bigr].
\label{eq:modal-power-cubic}
\end{equation}
These envelopes are generically essential-singular at $z=z_0-z_1$ when $\Lambda_{1,j}\ne0$.
The normal form in Section~\ref{sec:coord-transforms} belongs to the meromorphic subclass $d_4=0$, $\Lambda_{1,j}=0$.
There is therefore a distinction between imposing a cubic reduction on the original coefficients and reaching this more restricted cubic subclass by a transformation of a generic quartic trajectory.

\subsection*{Local singularity structure}
For nonsingular curves and generic, noncoincident arguments, the quartic envelopes have local exponents $r_{1,j}-1/2$ and $-r_{1,j}-1/2$ at $z=z_0-z_1$ and $z=z_0+z_1$, respectively, for $u_j$; the exponents are interchanged for $v_j$.
These follow from the equivalent local expression, with $x=z-z_0$ and $\nu_j=\mu_j-z_0$,
\begin{equation}
u_j(z)\propto e^{zr_{0,j}}\sigma(x+\nu_j)
\sigma(x+z_1)^{r_{1,j}-1/2}\sigma(x-z_1)^{-r_{1,j}-1/2}.
\end{equation}
The fractional powers still require compatible local branches.
Noninteger exponents give branch points; integer exponents can instead give zeros or poles.
In the cubic case, $\zeta(z-z_0+z_1)\sim1/(z-z_0+z_1)$ proves the essential singularity in \eqref{eq:u-v-cubic} when its coefficient is nonzero.
In contrast, the normal-form envelopes are meromorphic in the complex propagation variable.

% ---------- SECTION -------------
% --------------------------------

\section{Transforming to a normal form} \label{sec:coord-transforms}

In this section we present the second main result of this paper: a normal form for four-wave mixing that substantially simplifies both the governing equations and their solutions.
The structure of the analytic solutions in \eqref{eq:u-v-quartic} naturally suggests this transformation.
In particular, the presence of the factor $\sqrt{\wp(z_1) - \wp(z - z_0)}$ in the denominator motivates a local normalisation, while the $z$-dependent logarithmic terms weighted by $r_{1,j}$ suggest a gauge transformation to remove cross-phase modulation.

Although the transformation depends explicitly on the propagation variable $z$, it preserves the characteristic four-wave mixing interaction terms of the full coupled-mode equations while simplifying their coefficients.
In the normal-form coordinates, cross-phase modulation is eliminated, the differential equations for modal powers reduce from quartic to cubic polynomials, parameters are removed from the vector field through rescaling, entering instead through the initial data and reconstruction (Section~\ref{subsec:transform-domain}), and the solutions become single-valued meromorphic functions of Kronecker theta type.
The transformed systems retain Hamiltonian formulations and corresponding conserved intermodal power differences, with constants expressed in the transformed coordinates.
We accomplish this through a sequence of three transformations, which we now present in turn.

\subsection{Removing cross-phase modulation with a gauge transform}

To begin, let us consider a transform of the system in \eqref{eq:uv-system} of the following form:
\begin{align}
u_j(z) &= \hat{u}_j(z) e^{- \phi_j(z)}, \notag \\
v_j(z) &= \hat{v}_j(z) e^{\phi_j(z)}, \\
\sum_{j=1}^{4} \phi_j(z) &= 0,
\end{align}
where transforming the conjugate mode with the opposing phase leaves modal powers unchanged, and where the sum over $\phi_j$ being zero ensures we do not encounter any new terms appearing in the exponents of wave mixing terms.
We make the following choice for $\phi_j$ composed of a part linear in $z$ (labelled $\mathcal{L}$) and a part nonlinear in $z$ (labelled $\mathcal{NL}$) (where $\mathcal{L}$ and $\mathcal{NL}$ are function labels not exponents):
\begin{align}
\phi_j(z) &= \phi_j^{\mathcal{L}}(z) + \phi_j^{\mathcal{NL}}(z) \\
\phi_j^{\mathcal{L}}(z) &= z\left({a}_{j} - \frac{1}{4}\sum_{m=1}^{4} a_{m} - \frac{{\gamma}_{j}}{4} \sum\limits_{l,k=1}^{4} a_{l,k}\right) \\
\phi_j^{\mathcal{NL}}(z)  &= \sum_{k=1}^{4} \left({a}_{j,k} - \frac{1}{4}\sum_{l=1}^{4} {a}_{l,k}\right) \int_{0}^{z} \hat{u}_k (\xi) \hat{v}_k (\xi) \, d\xi \\
\sum_{j=1}^{4} \phi_j(z) &= \sum_{j=1}^{4} \phi_j^{\mathcal{L}}(z) = \sum_{j=1}^{4} \phi_j^{\mathcal{NL}}(z) =0.
\end{align}

This transforms \eqref{eq:uv-system} into the following system:
\begin{align}
\frac{\mathrm{d}}{\mathrm{d}z} \hat{u}_j &= \left(b_1 - 2b_2\hat{u}_j \hat{v}_j \right) \frac{\hat{u}_j}{4} + \prod\limits_{k=1, k \ne j}^{4} \hat{v}_k \notag \\
\frac{\mathrm{d}}{\mathrm{d}z} \hat{v}_j &= -\left(b_1 - 2b_2\hat{u}_j \hat{v}_j\right) \frac{\hat{v}_j}{4} - \prod\limits_{k=1, k \ne j}^{4} \hat{u}_k \label{eq:uv-hat-system}
\end{align}
for which the conserved canonical Hamiltonian is:
\begin{align}
\hat{H} &= \prod_{l=1}^{4} \hat{u}_l + \prod_{l=1}^{4} \hat{v}_l - \frac{1}{4}\sum_{l=1}^{4} \left(b_2 \hat{u}^{2}_l \hat{v}^{2}_l  - b_1 \hat{u}_l \hat{v}_l \right)
\end{align}
and the intermodal power conservation laws are unchanged such that $\hat{u}_j \hat{v}_j - \hat{u}_k \hat{v}_k = \gamma_j - \gamma_k$.
Comparing \eqref{eq:uv-hat-system} with the original system \eqref{eq:uv-system}, we observe significant simplification.
All modes now share a single propagation constant $b_1/4$ and a single self-phase modulation coefficient $2b_2/4$, whilst all cross-phase modulation terms $a_{j,k}$ with $j \ne k$ have been eliminated.
The coefficients $b_1$ and $b_2$ are fixed on the selected invariant set; $b_1$ depends on the conserved power differences, while $b_2$ depends only on the original system coefficients.
In terms of the solutions to the system, the effect of the $\phi_j^{\mathcal{NL}}$ transform is to remove the $z$ dependent log terms weighted by $r_{1,j}$ in \eqref{eq:u-v-quartic}.
This gauge removes accumulated mode-dependent diagonal terms; it is not an inversion of the full nonlinear propagation problem.

\subsection{Reducing the quartic to a cubic via local normalisation}\label{subsec:bar-coords}

The second transformation is a coordinate change involving a $z$-dependent normalisation of the modes, which induces a M\"obius transformation on the modal powers.
This implements the quartic to cubic reduction discussed in Section~\ref{sec:mode-power}, but here the effect is delivered at the level of the modes $u_j, v_j$ themselves rather than only at the modal power level $u_j v_j$, allowing us to observe corresponding changes in the coupled differential equations.
The transformation further reduces the number of parameters such that the propagation constant becomes the reciprocal of the SPM coefficient.

On the conserved power line, define $h(z)=\lambda_1d_5/(\lambda_1-\rho(z))$.
Its equivalent expressions in the transformed powers and the associated coordinate transform are $\left(\hat{u}_j, \hat{v}_j\right) \rightarrow \left(\bar{u}_j, \bar{v}_j \right)$ (note the hat to bar change in notation):
\begin{align}
\hat{u}_j(z) &= 2 \sqrt{{\lambda}_{1} \left({\gamma}_{j} - {\lambda}_{1}\right)} \frac{\bar{u}_j(z) e^{-\theta_j z}}{\sqrt{h{\left(z \right)}}} \notag \\
\hat{v}_j(z) &= 2 \sqrt{{\lambda}_{1} \left({\gamma}_{j} - {\lambda}_{1}\right)}  \frac{\bar{v}_j(z) e^{\theta_j z} }{\sqrt{h{\left(z \right)}}} \notag \\
h{\left(z \right)} &= {d}_{5} - \sum_{l=1}^{4} \left({\gamma}_{l} - {\lambda}_{1}\right) \bar{u}_l \bar{v}_l = 4 {\lambda}_{1} \bar{u}_j \bar{v}_j - \frac{{\lambda}_{1}{d}_{5} }{{\gamma}_{j} - {\lambda}_{1}} \label{eq:hat-to-bar}
\end{align}
where $\theta_j$ and $d_5$ are given in Appendix \ref{app:bar-coords-params}.
The $\theta_j$ parameter is not involved in the reduction of the quartic to the cubic but its inclusion keeps the propagation constant the same for all modes after the transform.
The key parameter in the transformation is the root $\lambda_1$ previously defined in Section \ref{sec:mode-power} for the purpose of quartic to cubic reduction.
The normalisation function $h(z)$ is constructed specifically to eliminate the factor $\sqrt{\wp(z_1) - \wp(z - z_0)}$ that appears in the denominator of \eqref{eq:u-v-quartic}.
We therefore already know its effect on the solutions through the simple application in \eqref{eq:hat-to-bar}, but it is revealing to observe the corresponding effect on the differential equations.

The transformation in \eqref{eq:hat-to-bar} sends \eqref{eq:uv-hat-system} to:
\begin{align}
\frac{\mathrm{d}}{\mathrm{d}z} \bar{u}_j &= -\left(\frac{1}{\chi} - \frac{\chi}{4}\bar{u}_j \bar{v}_j \right) \bar{u}_j - \frac{\varsigma \chi}{4} \prod\limits_{k=1, k \ne j}^{4} \bar{v}_k, \notag \\
\frac{\mathrm{d}}{\mathrm{d}z} \bar{v}_j &= \left( \frac{1}{\chi} - \frac{\chi}{4}\bar{u}_j \bar{v}_j\right) \bar{v}_j + \frac{\varsigma \chi}{4} \prod\limits_{k=1, k \ne j}^{4} \bar{u}_k \label{eq:uv-bar-system}
\end{align}
for which the conserved canonical Hamiltonian and intermodal power conservation laws are:
\begin{align}
\bar{H} &= \sum_{l=1}^{4} \left(\frac{\chi}{8} \bar{u}^{2}_l \bar{v}^{2}_l  - \frac{1}{\chi}  \bar{u}_l \bar{v}_l \right) - \frac{\varsigma \chi}{4} \left(\prod_{l=1}^{4} \bar{u}_l + \prod_{l=1}^{4} \bar{v}_l\right), \notag \\
\bar{\gamma}_j - \bar{\gamma}_k &= \bar{u}_j \bar{v}_j - \bar{u}_k \bar{v}_k,
\end{align}
with $\varsigma =\pm 1$, $\chi$ and $\bar{\gamma}_j$ given in Appendix \ref{app:bar-coords-params}.
The differences $\bar{\gamma}_j-\bar{\gamma}_k$ are constants of the transformed system and need not equal the original differences $\gamma_j-\gamma_k$.
The modal power in these coordinates evolves according to:
\begin{equation}
\label{eq:duv_bar_j}
\frac{d}{d z} \bar{u}_j \bar{v}_j = \frac{\varsigma \chi}{4}\left(\prod\limits_{k=1}^{4} \bar{u}_k - \prod\limits_{k=1}^{4} \bar{v}_k\right)
\end{equation}

Remarkably, the terms in \eqref{eq:uv-bar-system} are the same as those in \eqref{eq:uv-hat-system} only with different coefficients.
We call this a form-preserving invariance: coefficients change, while the characteristic full-mode interaction structure survives the explicit $z$ dependence of the normalisation.
Similarly, by comparing \eqref{eq:duv_bar_j} to \eqref{eq:duv_j} we see that \eqref{eq:duv_j} is invariant up to a constant factor.

To solve this system, we may define the function $\bar{w}(z)$ in terms of $h$ by:
\begin{align}
h{\left(z \right)} = \frac{\left({d}_{2} + 3 {d}_{3} {\lambda}_{1} + 6 {d}_{4} {\lambda}_{1}^{2}\right) {\lambda}_{1}}{3} - 4 \bar{w}{\left(z \right)} {\lambda}_{1}
\end{align}
and we find then that:
\begin{align}
\left(\frac{d}{d z} \bar{w}{\left(z \right)}\right)^{2} &= 4\,\bar{w}{\left(z \right)}^3 - g_2\,\bar{w}{\left(z \right)} - g_3, \notag \\
\bar{w}(z) &= \wp (z - z_0) \label{eq:dwp-bar} \\
h(z) &= 4\lambda_1\left(\wp (z_1) - \wp (z - z_0)\right) \label{eq:dwp-h-bar} \\
\bar{u}_j(z)\bar{v}_j(z) &= \wp(\mu_j - z_0) - \wp(z - z_0)
\end{align}
where $z_0$, $z_1$, $\mu_j$, $g_2$, and $g_3$ retain their previous definitions from the original coordinates.
We see that as a result of the transformation in \eqref{eq:hat-to-bar}, the right hand side of \eqref{eq:dwp-bar} is cubic in $\bar{w}$ and thus the corresponding differential equations for $h$ and $\bar{u}_j\bar{v}_j$ inherit this simplification as they are linearly related.
The elliptic function form of $h(z)$ in \eqref{eq:dwp-h-bar} confirms that it successfully eliminates the corresponding factor from the denominator in \eqref{eq:u-v-quartic}.

The solutions for modes $\bar{u}_j, \bar{v}_j$ are:
\begin{align}
\bar{u}_j(z) &= \frac{\bar{\alpha}_{j} \sigma{\left(z - 2 z_{0} + {\mu}_{j}\right)} \exp{\left( z \bar{\kappa}_j\right)} }{\sigma{\left(z - z_{0}\right)} \sigma{\left({\mu}_{j} - z_{0}\right)}} \notag \\
\bar{v}_j(z) &= \frac{\sigma{\left(z - {\mu}_{j}\right)} \exp{\left(- z \bar{\kappa}_j\right)}}{\bar{\alpha}_{j} \sigma{\left(z - z_{0}\right)} \sigma{\left({\mu}_{j} - z_{0}\right)} } \label{eq:u-v-bar}
\end{align}
where the Weierstrass $\sigma$ functions use the same $g_2,g_3$ from the original coordinates,
$\bar{\kappa}_j$ is given in Appendix \ref{app:bar-coords-params}, and $\bar{\alpha}_j$ is an integration constant fixed by initial conditions and used to capture any relative phase offset between a mode $\bar{u}_j$ and its conjugate $\bar{v}_j$.

\subsection{Why four modes are distinguished}\label{subsec:four-mode-invariance}

The local normalisation varies with the evolving modal powers and hence with propagation distance.
Nevertheless, its effect on the modal-power exchange equation is only a constant rescaling of the coupling coefficient.
The following calculation explains why four modes are distinguished within this class of square-root normalisations.
Consider $N$ unweighted mode pairs satisfying
\begin{equation}
I_j=u_jv_j=\gamma_j-\rho,\qquad
I_j'=\prod_{k=1}^{N}v_k-\prod_{k=1}^{N}u_k,
\label{eq:N-mode-power-exchange}
\end{equation}
where primes denote differentiation with respect to the original propagation variable $z$.
For constants $a,b$, define
\begin{equation}
h=(a+b\rho)^{-1},\qquad s_j^2=a+b\gamma_j,\qquad
u_j=\frac{s_jU_j}{\sqrt h},\quad
v_j=\frac{s_jV_j}{\sqrt h}.
\label{eq:N-mode-local-normalisation}
\end{equation}
We work locally where $h$ is finite and nonzero and all $s_j\ne0$, choosing consistent square-root branches.
The constants $s_j$ ensure that the new powers $J_j=U_jV_j=h(\gamma_j-\rho)/s_j^2$ have a common derivative:
\begin{align}
J_j'&=-h^2\rho' \notag\\
&=C\,h^{\,2-N/2}
\left(\prod_{k=1}^{N}V_k-\prod_{k=1}^{N}U_k\right),
\qquad C=\prod_{k=1}^{N}s_k.
\label{eq:N-mode-normalised-exchange}
\end{align}
Thus, for a nonconstant normalisation along an exchanging trajectory, the residual factor is constant if and only if $N=4$.
In that case the exchange equation retains its form in the same variable $z$, even though the normalisation depends on $z$; it behaves as though only a constant scaling had been applied.
For other $N$, removing the residual factor would require a nonconstant reparameterisation of $z$.
This calculation concerns the power equation; preservation of the full four-mode equation structure is established by \eqref{eq:uv-bar-system}.
At the scalar level, a M\"obius substitution $\rho=(m_{11}R+m_{12})/(m_{21}R+m_{22})$, with $\det\mathsf M\ne0$ for $\mathsf M=(m_{ij})_{i,j=1}^2$, sends $\rho'^2=\mathcal F(\rho)$ to
\begin{equation}
R'^2=\frac{(m_{21}R+m_{22})^4}{(\det\mathsf M)^2}\mathcal F\!\left(\frac{m_{11}R+m_{12}}{m_{21}R+m_{22}}\right).
\end{equation}
Thus the family of scalar polynomials of degree at most four is classically covariant; a cubic can also become a quartic.
The mode-level result is stronger than this scalar observation: the propagation-dependent square-root normalisation retains the complete four-mode equations and a Hamiltonian formulation, and acts on power exchange like constant scaling.
The cancellation above distinguishes four unweighted modes, independently of whether a lower-mode system might also have a quartic scalar reduction.
For weighted interaction products $\prod_j u_j^{\beta_j}$ and $\prod_j v_j^{\beta_j}$, the analogous factor is $h^{\,2-\frac12\sum_j\beta_j}$, so the distinguished condition is total interaction degree four, including degenerate four-wave mixing with weights $(1,1,2)$.

\subsection{A parameterless system via rescaling}\label{subsec:tilde-coords}

The final transformation is a simple rescaling that eliminates all remaining parameters from the system.
We scale the modes and the propagation variable as follows (note the bar to tilde notation change):
\begin{align}
\bar{u}_j(z) &= \frac{\varsigma \sqrt[4]{\varsigma}}{\chi}\tilde{u}_j(\xi), \notag \\
\bar{v}_j(z) &= \frac{\varsigma }{\sqrt[4]{\varsigma}\,\chi}\tilde{v}_j(\xi), \notag \\
z &= \chi\xi
\end{align}
where the specific form of the mode scaling is chosen to yield parameterless coefficients in the transformed equations (whilst retaining a coefficient of 4 in the cubic term of the corresponding differential equations as per the Weierstrass form).
This substitution transforms \eqref{eq:uv-bar-system} to:
\begin{align}
\frac{\mathrm{d}}{\mathrm{d}\xi} \tilde{u}_j &= -\left(1 - \frac{\tilde{u}_j \tilde{v}_j}{4} \right) \tilde{u}_j - \frac{1}{4}\prod\limits_{k=1, k \ne j}^{4} \tilde{v}_k, \notag \\
\frac{\mathrm{d}}{\mathrm{d}\xi} \tilde{v}_j &= \left( 1 - \frac{\tilde{u}_j \tilde{v}_j}{4}\right) \tilde{v}_j + \frac{1}{4}\prod\limits_{k=1, k \ne j}^{4} \tilde{u}_k, \label{eq:uv-tilde-system}
\end{align}

for which the conserved canonical Hamiltonian is:
\begin{align}
\tilde{H} &= \sum_{l=1}^{4} \left(\frac{1}{8} \tilde{u}^{2}_l \tilde{v}^{2}_l  - \tilde{u}_l \tilde{v}_l \right) - \frac{1}{4}\left(\prod_{l=1}^{4} \tilde{u}_l + \prod_{l=1}^{4} \tilde{v}_l\right), \label{eq:H-tilde}
\end{align}

and the solutions are:
\begin{align}
\tilde{u}_j{\left(\xi\right)} &= \frac{\tilde{\alpha}_{j} \,\sigma{\left(\xi - 2 \xi_{0} + {\tilde{\mu}}_{j}\right)} \exp\left({\xi \left( 1 + \frac{{\tilde{\gamma}}_{j}}{2} - \zeta{\left( {\tilde{\mu}}_{j} - \xi_{0}\right)}\right)}\right)}{\sigma{\left(\xi - \xi_{0}\right)} \sigma{\left( {\tilde{\mu}}_{j} - \xi_{0}\right)}}, \notag \\
\tilde{v}_j{\left(\xi\right)} &= \frac{ \sigma{\left(\xi - {\tilde{\mu}}_{j}\right)} \exp\left({-\xi \left( 1 + \frac{{\tilde{\gamma}}_{j}}{2} - \zeta{\left( {\tilde{\mu}}_{j} - \xi_{0}\right)}\right)}\right)}{\tilde{\alpha}_{j} \,\sigma{\left(\xi - \xi_{0}\right)} \sigma{\left( {\tilde{\mu}}_{j} - \xi_{0}\right)}}  \label{eq:u-v-tilde}
\end{align}
where the $\sigma$ and $\zeta$ functions use the scaled invariants $\tilde{g}_2$ and $\tilde{g}_3$ given in Appendix \ref{app:scaled-params} together with $\xi_0$, $\tilde{\alpha}_j$, $\tilde{\mu}_j$ and $\tilde{\gamma}_j$.

The solutions in both \eqref{eq:u-v-bar} and \eqref{eq:u-v-tilde} are single-valued meromorphic functions of the propagation variable, expressed as shifted $\sigma$ ratios multiplied by exponential factors.
More precisely, they are exponentially gauged Kronecker kernels, or equivalently genus-one Baker--Akhiezer functions with elementary exponential factors.
The relation between these conventions, and its connection to higher-genus systems, is discussed in Section~\ref{subsec:kronecker-ba}.

Owing to the parameter-free form of \eqref{eq:uv-tilde-system}, the simple structure of the Hamiltonian \eqref{eq:H-tilde}, and the elegant Kronecker theta function solutions in \eqref{eq:u-v-tilde}, we refer to \eqref{eq:uv-tilde-system} as the normal form of four-wave mixing.
The derivation of this normal form, together with the local transformation from the generic quartic system on the domain specified below, constitutes the second main result of this paper.

\subsection{Domain and inversion of the transformations}
\label{subsec:transform-domain}

The construction is local on a fixed set of conserved quantities.
Here ``normal form'' refers to a universal target system for trajectories: the original parameters enter its initial data and the reconstruction, and the transformation uses the quadrature $\int\rho\,dz$.
Assume a nonsingular elliptic curve and choose a simple quartic root $\lambda_1$ such that
\begin{equation}
\lambda_1\ne0,\qquad d_5=\mathcal F'(\lambda_1)\ne0,\qquad
q_1=Q(\lambda_1)\ne0.
\end{equation}
The root relation $q_1^2=4\prod_j(\gamma_j-\lambda_1)$ then ensures $\gamma_j-\lambda_1\ne0$ and $\chi=8q_1/d_5\ne0$.
On the conserved power line, \eqref{eq:hat-to-bar} is equivalently
\begin{equation}
h(z)=\frac{\lambda_1d_5}{\lambda_1-\rho(z)},\qquad
\bar u_j=\frac{\sqrt h\,e^{\theta_jz}}{c_j}\hat u_j,\qquad
\bar v_j=\frac{\sqrt h\,e^{-\theta_jz}}{c_j}\hat v_j,
\quad c_j^2=4\lambda_1(\gamma_j-\lambda_1).
\label{eq:local-inverse-normalisation}
\end{equation}
Choose the constant roots consistently: set $S=\prod_jc_j/(16\lambda_1^2)$ and $\varsigma=q_1/(2S)=\pm1$.
This fixes the square root in the definition of $\varsigma$ in Appendix~\ref{app:bar-coords-params}.
Choose a continuous branch of $\sqrt h$ on a simply connected region where $h$ is finite and nonzero, excluding poles and $\rho=\lambda_1$.
The inverse normalisation is \eqref{eq:hat-to-bar}; the first gauge is inverted using its stated integrals of the unchanged modal products, and the final rescaling is invertible since $\chi\ne0$.
The fourth root of $\varsigma$ in that rescaling is also fixed consistently.
These are transformations of solutions on the chosen invariant set; the Hamiltonian formulations of the resulting equations do not assert that the full transformation preserves the original symplectic form.
The condition $\lambda_1\ne0$ is specific to this chosen normalisation, rather than an intrinsic obstruction.
If a selected root is reached on the propagation path, the normalisation has a pole; for a whole bounded physical oscillation, choose an admissible root not visited by that orbit when available, or use overlapping local charts.
Zeros of $h$ at $z_0\pm z_1$ account for the common square-root part of the original branch behaviour; the preceding gauge removes the mode-dependent part.
Writing $r^4=\varsigma$, the composite multipliers are
\begin{equation}
F_j(z)=\frac{\chi\sqrt h}{\varsigma r c_j}e^{\theta_jz+\phi_j(z)},\qquad
G_j(z)=\frac{\chi r\sqrt h}{\varsigma c_j}e^{-\theta_jz-\phi_j(z)}.
\end{equation}
Thus $\tilde u_j=F_j(z)u_j$, $\tilde v_j=G_j(z)v_j$, and the physical real slice becomes
\begin{equation}
\tilde v_j=-is_j\frac{G_j(z)}{\overline{F_j(z)}}\overline{\tilde u_j}.
\end{equation}
This transported constraint is generally propagation dependent; physical fields are recovered by the inverse transformation.
If $\chi$ is complex, real physical $z$ corresponds to a line $\xi=z/\chi$ in the complex $\xi$ plane.
Exceptional roots, singular curves, and zeros of the chart denominators require another chart or a separate limiting treatment.

% ---------- SECTION -------------
% --------------------------------

\section{Numerical validation and implementation} \label{sec:num-val}

In this section we validate the analytic solutions through numerical evaluation and demonstrate their practical implementation using open-source software.
This serves two purposes: first, to verify that the solutions correctly satisfy the differential equations, and second, to show that the Weierstrass elliptic functions can be readily evaluated using standard numerical libraries.

We evaluate the analytic solutions using the notebook's wrapper based on the \emph{pyweierstrass} Python package \cite{laurentPyweierstrass2022}, which provides Weierstrass elliptic functions as a wrapper around Jacobi theta functions from the \emph{mpmath} package \cite{MpmathPythonLibrary2023}.
For comparison, we solve the differential equations numerically using the DOP853 Runge-Kutta algorithm from \emph{SciPy} \cite{virtanenSciPy10Fundamental2020}.
For the quartic formulas, the logarithm is continued along the sampled propagation path by phase unwrapping, with the branch fixed at the initial point.

For plotting purposes we define the following complex-valued functions $\mathcal{P}_j(z),\ \phi_j(z) \in \mathbb{C}$ that are analogous to modal power and phase; they would be equivalent to the corresponding real-valued functions if $v_j$ were the complex conjugate of $u_j$ rather than the more general Hamiltonian conjugate that we typically consider in our generalised systems:
\begin{align}
\mathcal{P}_j(z) &= u_j(z)v_j(z), \notag \\
\phi_j(z) &= \frac{\log\left(\frac{u_j(z)}{v_j(z)}\right)}{2 i}. \label{eq:A-phi-u-v}
\end{align}

\begin{figure}[H]
\centering
\includegraphics[width=1.0\textwidth]{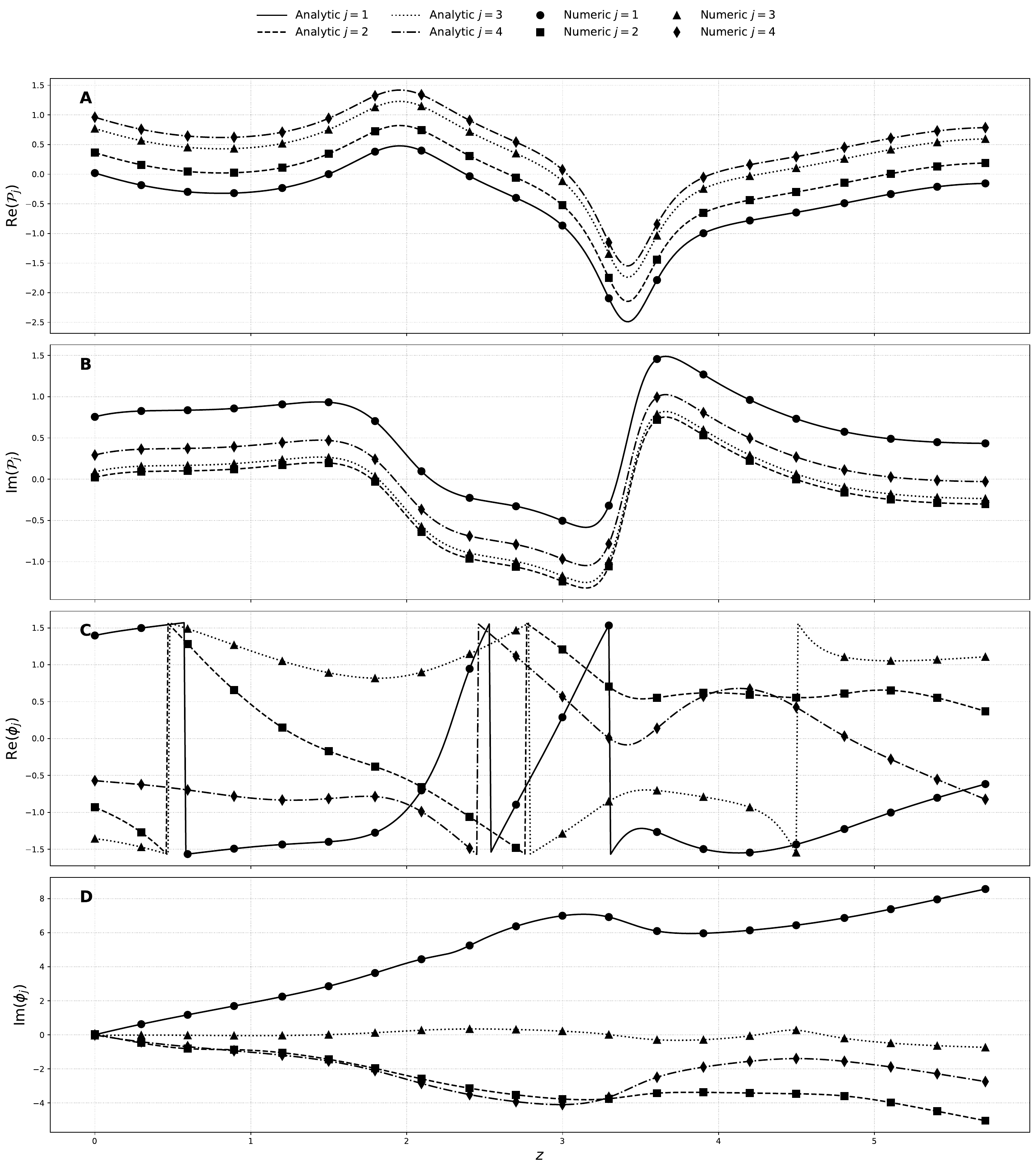}
\caption{\textbf{A} real and \textbf{B} imaginary parts of the complex mode power $\mathcal{P}_j=u_j v_j$, and \textbf{C} real and \textbf{D} imaginary parts of the complex phase variable $\phi_j$, using the analytic solutions in \eqref{eq:u-v-quartic} (dashed lines) compared against numerical integration of the abstract system \eqref{eq:uv-system} (symbols).}
\label{fig:case1-plot1}
\end{figure}

The top row of Figure \ref{fig:case1-plot1} shows the real, \textbf{A}, and imaginary, \textbf{B}, parts of the complex mode power $\mathcal{P}_j$,
while the bottom row shows the real, \textbf{C}, and imaginary, \textbf{D}, parts of the complex phase variable $\phi_j$, using the analytic solutions in \eqref{eq:u-v-quartic} (dashed lines) compared against numerical integration of the abstract system \eqref{eq:uv-system} (symbols).
The intermodal power conservation laws in \eqref{eq:power-conserved} are evident from the coordinated evolution of all four modes.
The initial conditions and parameter values used to generate Figure \ref{fig:case1-plot1} are listed in Appendix~\ref{app:case1-params}, Tables~\ref{tab:case1-init} and \ref{tab:case1-params}.
For the dimensionless parameter set shown in Figure~\ref{fig:case1-plot1}, the maximum componentwise absolute discrepancy between the analytic eight-component solution vector $(u_1,\ldots,u_4,v_1,\ldots,v_4)$ and the DOP853 numerical solution, over the plotted interval and sampled points, was approximately $8\times10^{-10}$.
Dividing this discrepancy by the maximum absolute component of the numerical solution gives approximately $2\times10^{-13}$; this is a scale-normalised error, rather than a pointwise relative error.

\begin{figure}[H]
\centering
\includegraphics[width=1.0\textwidth]{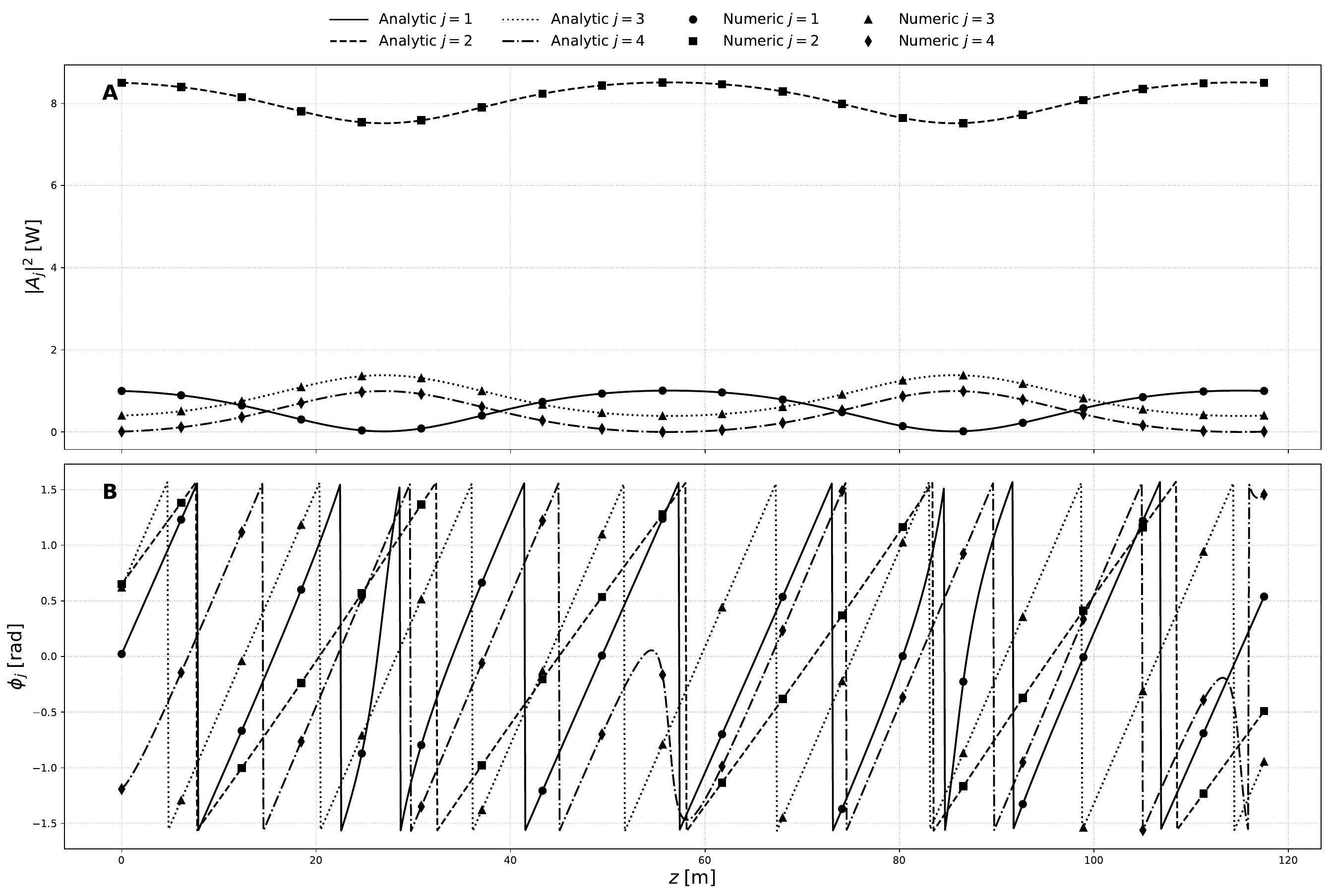}
\caption{\textbf{A} Intensity $\mathcal{P}_j=|A_j|^2$, and \textbf{B} phase $\phi_j$ of physical field amplitudes: analytic solution \eqref{eq:u-v-quartic} converted via \eqref{eq:u-v-A} (dashed lines) compared with numerical integration (symbols).}
\label{fig:case2-plot1}
\end{figure}

Figure \ref{fig:case2-plot1} shows \textbf{A} (top row) the intensity ($\mathcal{P}_j=\left|A_j\right|^2$), and \textbf{B} (bottom row) phase evolution of the physical field amplitudes $A_j$ in a test case for the physical four-wave mixing equations.
In this case $A_j, A^*_j$ are used in place of $u_j, v_j$ respectively in \eqref{eq:A-phi-u-v}.
The analytic solutions are plotted using \eqref{eq:u-v-quartic} to calculate $u_j, v_j$, then converting to $A_j, A_j^{*}$ via \eqref{eq:u-v-A} (dashed lines), and are compared against numerical integration of the original physical FWM system \eqref{eq:fwm-system} (symbols).
Figure \ref{fig:case2-plot1} \textbf{A} shows the periodic energy exchange between the waves, while \textbf{B} compares their phase evolution using the same logarithmic phase convention for both solutions.
The plot spans two power-exchange periods, approximately $117.5\,\mathrm m$.
This illustrative scalar-fibre example uses powers $(1,8.5,0.4,0.01)$ W, a common effective area $A_{\rm eff}=10\,\mu\mathrm m^2$, and $\beta^{(2)}=-30\,\mathrm{ps}^2/\mathrm{km}$.
With frequency offsets $(-\Omega,\Omega,-3\Omega,3\Omega)$, $\Omega=2\pi\times100$ GHz, it has $\Delta k=8\beta^{(2)}\Omega^2\simeq-9.47\times10^{-2}\,\mathrm m^{-1}$.
The frequency-independent and frequency-linear carrier terms have been removed from $\beta_j$; propagation distance is in metres and powers are in watts.
All overlaps use the common-transverse-mode approximation $f_{j,k}=f_{1,2,3,4}=A_{\rm eff}^{-1}$ and have units $\mathrm m^{-2}$.
The initial conditions and physical parameter values used to generate Figure \ref{fig:case2-plot1} are provided in Appendix~\ref{app:case2-params}, Tables~\ref{tab:case2-init} and \ref{tab:case2-physical}.
For this example, the maximum absolute discrepancy in the reconstructed complex physical fields, divided by the maximum field amplitude over the sampled interval, is approximately $2\times10^{-11}$.

The analytic and numeric solutions are in excellent agreement in all plots.
These comparisons provide a numerical check of the analytic solutions and demonstrate the feasibility of their numerical evaluation using readily available software tools.
The accompanying notebook includes an additional comparison for the $d_4=0$ cubic formulas in a two-field reduction with $\Lambda_{1,j}=0$.
A separate reproducible four-mode check uses $b_2=\pm2$ with all $\Lambda_{1,j}\ne0$, checking both the zeta exponential and the linear coefficient in \eqref{eq:u-v-cubic} against the mode equations and direct integration.
The supplied audit scripts also verify the full gauge/normalisation/rescaling chain for both coupling signs and check all eight normal-form envelopes for three complex initial states; their equation residuals are below $10^{-58}$ at 60-digit working precision.

\section{Discussion}

\subsection{Kronecker kernels and Baker--Akhiezer functions}
\label{subsec:kronecker-ba}

The functions appearing in the normal-form solutions are used under different names and normalisations in the literature.
To make the correspondence explicit, let $\vartheta(s;\tau)$ denote the odd Jacobi theta function with quasiperiods $1$ and $\tau$.
The Kronecker function is
\begin{equation}
F(s,t;\tau)=
\frac{\vartheta'(0;\tau)\vartheta(s+t;\tau)}
{\vartheta(s;\tau)\vartheta(t;\tau)},
\label{eq:kronecker-kernel}
\end{equation}
where the prime denotes differentiation in the first argument; see \cite[Section~3.4]{brownMultipleEllipticPolylogarithms2013}.
For the period convention used here, $\tau=\omega_3/\omega_1$ and $\eta_1=\zeta(\omega_1)$, define
\begin{align}
\Phi(x,a)
&=\frac{\sigma(x+a)}{\sigma(x)\sigma(a)}\exp[-x\zeta(a)] \notag\\
&=\frac{1}{2\omega_1}
\exp\!\left[x\left(\frac{\eta_1a}{\omega_1}-\zeta(a)\right)\right]
F\!\left(\frac{x}{2\omega_1},\frac{a}{2\omega_1};\tau\right).
\label{eq:ba-kronecker}
\end{align}
Thus the Kronecker and Baker--Akhiezer descriptions involve an explicit exponential change of normalisation.
We use the plus sign in $\sigma(x+a)$; Krichever's elliptic Baker--Akhiezer function uses the equivalent convention $a\mapsto-a$ \cite[Eq.~(1.7)]{kricheverEllipticSolutionsNonlinear1994}.
For fixed $a$ outside the period lattice, $\Phi$ is meromorphic and quasiperiodic in $x$, rather than doubly periodic.
In its spectral argument $a$, it has the prescribed exponential singularity at the origin of the elliptic curve.
This illustrates the defining Baker--Akhiezer construction: meromorphic data away from marked points, together with prescribed exponential behaviour at those points.

For example, set $x=\xi-\xi_0$, $a_j=\tilde\mu_j-\xi_0$, $c_j=1+\tilde\gamma_j/2$, and $B_j=\tilde\alpha_j\exp[-\xi_0\zeta(a_j)]$.
Remove the reciprocal exponential factors by defining the gauged envelopes
\begin{align}
U_j(x)&=e^{-c_j\xi}\tilde u_j(\xi)=B_j\Phi(x,a_j),\notag\\
V_j(x)&=e^{c_j\xi}\tilde v_j(\xi)=-B_j^{-1}\Phi(x,-a_j),
\qquad x=\xi-\xi_0.
\label{eq:tilde-ba-pair}
\end{align}
The minus sign follows from the oddness of $\sigma$.
The normal-form equations \eqref{eq:uv-tilde-system} give the Wronskian directly:
\begin{equation}
\mathcal W_j:=U_j'V_j-U_jV_j'
=\tilde H-\frac18\sum_{k=1}^4\tilde\gamma_k^2
-\frac12\tilde\gamma_j^2-4\tilde\gamma_j
=\wp'(a_j).
\label{eq:ba-wronskian}
\end{equation}
Here primes denote $d/dx$; the right-hand side is constant because $\tilde H$ and the $\tilde\gamma_k$ are conserved.
Differentiating this Wronskian gives $U_j''V_j-U_jV_j''=0$, so the two envelopes share a second-order potential wherever their product is nonzero.
Direct substitution of the normal-form equations gives that potential explicitly:
\begin{align}
U_j''&=\mathcal V_j(x)U_j,\qquad
V_j''=\mathcal V_j(x)V_j,\notag\\
\mathcal V_j(x)&=2\wp(x)+\wp(a_j)\notag\\
&=4-\frac{\tilde H}{4}+\frac1{16}\sum_{k=1}^4\tilde\gamma_k^2
+\tilde\gamma_j-\frac12\sum_{k=1}^4 U_k(x)V_k(x).
\label{eq:ba-lame}
\end{align}
Thus the propagation-dependent part of the potential is generated by the modal products themselves, with mode-dependent constant spectral shifts.
In elliptic notation this is the one-gap Lam\'e equation \cite[Eq.~(1.9)]{kricheverEllipticSolutionsNonlinear1994}; its solutions here are specifically $U_j(x)$ and $V_j(x)$.
These identities extend meromorphically across isolated zeros.
This describes the exponentially gauged envelopes without changing the autonomous normal form \eqref{eq:uv-tilde-system}.

The Kronecker description connects these solutions to classical Fourier series \cite{whittakerCourseModernAnalysis2021} and multipole expansions \cite{weilEllipticFunctionsAccording1976,charolloisEllipticFunctionsAccording2016}, while the Baker--Akhiezer description places them alongside the spectral functions of integrable systems.
Related kernels also occur in two-wave and three-wave mixing in quadratic media, polarisation dynamics in nonlinear fibres, and parity-time symmetric couplers \cite{heskethGeneralComplexEnvelope2015}.
These connections concern the shared analytic structure of the solutions; they do not require the underlying optical models to be identical.

The Weierstrass notation also makes the passage to hyperelliptic sigma functions transparent.
For example, the genus-two Baker function is a shifted sigma ratio with an exponential involving second-kind Abelian integrals \cite[Eq.~(3.3)]{bradenBilinearRecurrencesAddition2005}, generalising the $\zeta(a)$ term in \eqref{eq:ba-kronecker}.
The author's related work \cite{zotero-item-855} extends the polynomial coupled-mode Hamiltonian setting to smooth odd-degree hyperelliptic curves, constructing local mode solutions from sigma functions and Abelian integrals and relating their conservation laws to higher-genus determinant identities.
The genus-one calculation here therefore provides a concrete model for both the function-theoretic and determinant aspects of that extension.
The higher-genus reconstructed modes should, however, be distinguished from an established Baker--Akhiezer spectral construction: their global continuation and the associated spectral problem require further analysis.

\section{Conclusion}

We have presented analytic solutions for the full complex envelopes of the quasi-continuous-wave four-wave mixing model considered here in the generic quartic regime.
The solutions, expressed in terms of Weierstrass elliptic functions, provide the full complex field envelopes for all four waves under generic initial conditions, including pump depletion, cross-phase modulation, and phase mismatch effects.
These solutions complement previous work in the literature by providing a unified treatment of the complex envelopes themselves rather than treating amplitudes and phases separately.

Guided by the structure of these solutions, we developed a sequence of coordinate transformations that reveal a normal form of the four-wave mixing equations.
In these normal-form coordinates, the system becomes parameter-free on the local transformation domain stated in Section~\ref{subsec:transform-domain}.
The solutions in normal-form coordinates are single-valued meromorphic functions of Kronecker theta type, connecting four-wave mixing to a broader class of integrable nonlinear optical systems.
The transformations preserve the essential structure of the differential equations despite depending explicitly on the propagation variable, exhibiting a mode-level form-preserving invariance whose interaction-degree-four cancellation goes beyond the scalar quartic-to-cubic reduction.

The connection between the Hamiltonian formulation and the Frobenius-Stickelberger determinant, established in Section~\ref{sec:fs-det-ham-proof}, represents the conserved Hamiltonian relation in elliptic-function terms and links four-wave mixing to classical addition identities.
This mathematical framework may prove valuable for analysing related parametric processes and could extend to higher-order mixing or multimode interactions.

The methods developed here are broadly applicable to related nonlinear systems.
In particular, the approach may be transferable to the analysis of mode coupling in multimode nonlinear fibre optics (see, e.g., Chapter 4 in \cite{heskethNonlinearEffectsMultimode2014}), where similar coupled-mode structures arise.

Numerical evaluation using open-source Python libraries confirms the analytic expressions against direct integration and illustrates their computational accessibility.
These results describe the elliptic structure of the stated continuous-wave coupled-wave approximation and its local normal-form reconstruction.

\section*{Acknowledgements}
The author thanks Maria Dimou for a careful read of the manuscript.

\section*{Code Availability}
The code used in this paper is available at \cite{heskethHeskethGDFwmpaper2026} and includes Python Jupyter notebooks with symbolic derivations using \emph{SymPy} \cite{meurerSymPySymbolicComputing2017} as well as the code for numerical validation used in Section \ref{sec:num-val}.

\appendix

% ---------- APPENDIX ------------
% --------------------------------

\section{Parameter definitions in the original coordinates}
\label{app:param-def-appendix}

This appendix provides explicit formulae for the parameters $b_l$, $c_l$, and $d_l$ that appear in the quartic polynomial \eqref{eq:drho-sqrd-3} and related expressions throughout Section~\ref{sec:mode-power}.
These parameters are expressed in terms of the system parameters $a_j$, $a_{j,k}$ and the conserved constants $\gamma_j$.

\begin{align}
{b}_{0} &= {a}_{0} + \sum_{j=1}^{4} {a}_{j} {\gamma}_{j} + \frac{1}{2}\sum\limits_{j,k=1}^{4} {a}_{j,k} {\gamma}_{j} {\gamma}_{k}, \notag \\
{b}_{1} &= - \sum_{j=1}^{4} {a}_{j} - \frac{1}{2}\sum\limits_{j,k=1}^{4} \left({\gamma}_{j} + {\gamma}_{k}\right) {a}_{j,k}, \notag \\
{b}_{2} &= \frac{1}{2}\sum\limits_{j,k=1}^{4} {a}_{j,k}, \notag \\
{c}_{0} &= \prod_{j=1}^{4} {\gamma}_{j}, \notag \\
{c}_{1} &= - \sum_{k=1}^{4} \prod_{\substack{j=1 \\ j \ne k}}^{4} {\gamma}_{j}, \notag \\
{c}_{2} &= \frac{3}{2} \sum_{j=1}^{4} {\gamma}_{j}^{2} - \frac{1}{4}\sum\limits_{j,k=1}^{4} \left({\gamma}_{j} - {\gamma}_{k}\right)^{2} = - \frac{1}{2}\sum_{j=1}^{4} {\gamma}_{j}^{2}, \notag \\
{d}_{0} &= {b}_{0}^{2} - 4 {c}_{0}, \notag \\
{d}_{1} &= 2 {b}_{0} {b}_{1} - 4 {c}_{1}, \notag \\
{d}_{2} &= 2 {b}_{0} {b}_{2} + {b}_{1}^{2} - 4 {c}_{2}, \notag \\
{d}_{3} &= 2 {b}_{1} {b}_{2}, \notag \\
{d}_{4} &= {b}_{2}^{2} - 4
\end{align}

% ---------- APPENDIX ------------
% --------------------------------

\section{Initial condition relations}
\label{app:init-conds}

This appendix gives the relationships between initial conditions for the normalised system and the constants $\gamma_j$ and $\rho(0)$ used throughout the analysis.
These relations enable determination of the constants from specified initial values of $u_j(0)$ and $v_j(0)$.

\begin{align}
\rho{\left(0 \right)} &= - \frac{1}{4}\sum_{j=1}^{4} u_j{\left(0 \right)} v_j{\left(0 \right)}, \notag \\
\left. \frac{d}{d z} \rho{\left(z \right)} \right|_{\substack{ z=0 }} &= \prod_{j=1}^{4} u_j(0) - \prod_{j=1}^{4} v_j(0), \notag \\
{\gamma}_{j} &= u_j{\left(0\right)} v_j{\left(0 \right)} + \rho{\left(0 \right)}
\end{align}

% ---------- APPENDIX ------------
% --------------------------------

\section{Parameter definitions in the bar coordinates}
\label{app:bar-coords-params}

This appendix provides the bar-coordinate parameters.
The square root $S=\sqrt{\prod_j(\gamma_j-\lambda_1)}$ below is chosen as $S=\prod_j c_j/(16\lambda_1^2)$ with the same constant mode roots $c_j^2=4\lambda_1(\gamma_j-\lambda_1)$ used in Section~\ref{subsec:transform-domain}; hence $\varsigma=Q(\lambda_1)/(2S)$.

\begin{align}
{d}_{5} &= {d}_{1} + 2 {d}_{2} {\lambda}_{1} + 3 {d}_{3} {\lambda}_{1}^{2} + 4 {d}_{4} {\lambda}_{1}^{3}, \notag  \\
\varsigma & = \frac{1}{2} \left(b_0 + b_1 {\lambda}_{1} + b_2{\lambda}_{1}^2\right) \left(\sqrt{\prod\limits_{j=1}^4 \left(\gamma_j - \lambda_1\right)}\right)^{-1}  = \pm 1, \\
\chi &= \frac{8\left(b_0 + b_1 {\lambda}_{1} + b_2{\lambda}_{1}^2\right)}{d_5} = \frac{16\, \varsigma}{d_5}\sqrt{\prod\limits_{j=1}^4 \left(\gamma_j - \lambda_1\right)}, \notag \\
\bar{\gamma}_{j} &= \frac{{d}_{5}}{4 \left({\gamma}_{j} - {\lambda}_{1}\right)} + \frac{{b}_{1} + 2 {b}_{2} {\lambda}_{1}}{\chi} - \frac{4}{\chi^{2}}, \notag \\
{\theta}_{j} &= \frac{b_0 + b_1 {\lambda}_{1} + b_2{\lambda}_{1}^2}{2\left({\gamma}_{j} - {\lambda}_{1}\right)} + \frac{{b}_{1}}{4} + \frac{{b}_{2} {\gamma}_{j}}{2} + \frac{{b}_{2} {\lambda}_{1}}{2} - \frac{1}{\chi}, \quad \sum\limits_{j=1}^{4} \theta_j = 0, \notag \\
\bar{\kappa}_j &= 2 {\theta}_{j} + \frac{1}{\chi} - {b}_{2} {\gamma}_{j} - \zeta{\left({\mu}_{j} - z_{0}, g_2, g_3\right)}.
\end{align}

% ---------- APPENDIX ------------
% --------------------------------

\section{Scaled parameters in the tilde coordinates}
\label{app:scaled-params}
When the system in subsection \ref{subsec:tilde-coords} (tilde notation) is considered a transform of that in subsection \ref{subsec:bar-coords} (bar notation), then parameters are related via the following scaling laws:
\begin{align}
\tilde{H} &= \chi^3\bar{H}, \\
\tilde{\alpha}_j &= \frac{\varsigma}{\sqrt[4]{\varsigma}}\,\bar{\alpha}_j, \notag \\
\tilde{\gamma}_j &= \chi^2\bar{\gamma}_j, \notag \\
\tilde{g}_2 &= \chi^4 g_2, \notag \\
\tilde{g}_3 &= \chi^6 g_3, \notag \\
\xi_0 &= \frac{z_0}{\chi}, \notag \\
\tilde{\mu}_j &= \frac{\mu_j}{\chi}.
\end{align}

If the system in subsection \ref{subsec:tilde-coords} is considered as a standalone system irrespective of any transformation,
then the parameters can be given exclusively in the coordinate system denoted by tilde notation as follows.
Firstly, $\tilde{H}$ can be obtained from substituting $\tilde{u}_j{\left(0\right)}, \tilde{v}_j{\left(0 \right)}$ into \eqref{eq:H-tilde}
and $\tilde{\alpha}_j$ can be considered an integration constant fixed by initial conditions to capture any phase offset between $\tilde{u}_j, \tilde{v}_j$.
We may further use relations:
\begin{align}
\tilde{\gamma}_{j} =& \tilde{u}_j{\left(0\right)} \tilde{v}_j{\left(0 \right)} - \frac{1}{4}\sum_{j=1}^{4} \tilde{u}_j{\left(0\right)} \tilde{v}_j{\left(0\right)}, \notag \\
\tilde{g}_{2} =& \frac{\tilde{H}^{2}}{12} + \tilde{H} \left(\frac{{\tilde{e}}_{2}}{12} + \frac{16}{3}\right) + \frac{{\tilde{e}}_{2}^{2}}{48} + \frac{2 {\tilde{e}}_{2}}{3} - \frac{{\tilde{e}}_{3}}{4} + \frac{64}{3}, \notag \\
\tilde{g}_{3} =& \frac{\tilde{H}^{3}}{216} + \tilde{H}^{2} \left(\frac{{\tilde{e}}_{2}}{144} - \frac{5}{9}\right) + \tilde{H} \left(\frac{{\tilde{e}}_{2}^{2}}{288} - \frac{2 {\tilde{e}}_{2}}{9} - \frac{{\tilde{e}}_{3}}{48} - \frac{64}{9}\right) \notag \\
&+ \frac{{\tilde{e}}_{2}^{3}}{1728} - \frac{5 {\tilde{e}}_{2}^{2}}{144} - \frac{{\tilde{e}}_{2} {\tilde{e}}_{3}}{96} - \frac{8 {\tilde{e}}_{2}}{9} + \frac{{\tilde{e}}_{3}}{3} + \frac{{\tilde{e}}_{4}}{4} - \frac{512}{27}
\end{align}
where $\tilde{e}_j=\tilde{e}_j\left(\tilde{\gamma}_1, \tilde{\gamma}_2, \tilde{\gamma}_3, \tilde{\gamma}_4\right)$ is the $j^{th}$ order \emph{elementary symmetric polynomial} formed with the four $\tilde{\gamma}_j$. The points $\xi_0, \tilde{\mu}_j$ are found by inverting:
\begin{align}
\wp{\left(\xi_{0},\tilde{g}_{2},\tilde{g}_{3} \right)} &= - \frac{\tilde{H}}{12} - \frac{{\tilde{e}}_{2}}{24} - \frac{\sum_{j=1}^{4} \tilde{u}_j{\left(0\right)} \tilde{v}_j{\left(0\right)}}{4} + \frac{4}{3}, \notag \\
\wp'{\left(\xi_{0},\tilde{g}_{2},\tilde{g}_{3} \right)} &= \frac{1}{4}\left(\prod_{j=1}^{4}{\tilde{u}_j(0)} - \prod_{j=1}^{4}{\tilde{v}_j(0)}\right), \notag \\
\wp{\left({\tilde{\mu}}_{j} - \xi_{0},\tilde{g}_{2},\tilde{g}_{3} \right)} &= - \frac{\tilde{H}}{12} - \frac{{\tilde{e}}_{2}}{24} + {\tilde{\gamma}}_{j} + \frac{4}{3}, \notag \\
\wp'{\left({\tilde{\mu}}_{j} - \xi_{0},\tilde{g}_{2},\tilde{g}_{3} \right)} &= \tilde{H} + \frac{{\tilde{e}}_{2}}{4} - \frac{{\tilde{\gamma}}_{j}^{2}}{2} - 4 {\tilde{\gamma}}_{j}.
\end{align}

% ---------- APPENDIX ------------
% --------------------------------

\section{Parameters for Figure \ref{fig:case1-plot1}}
\label{app:case1-params}

\noindent
\begin{minipage}[t]{0.48\textwidth}
\centering
\captionof{table}{Initial mode values in Figure \ref{fig:case1-plot1}.}
\label{tab:case1-init}
\begin{tabular}{c c}
\hline
Variable & Initial value \\
\hline
$u_1(0)$ & $0.486-0.712i$ \\
$u_2(0)$ & $0.358-0.454i$ \\
$u_3(0)$ & $0.243-0.877i$ \\
$u_4(0)$ & $-0.918+0.413i$ \\
$v_1(0)$ & $-0.711+0.512i$ \\
$v_2(0)$ & $0.36+0.514i$ \\
$v_3(0)$ & $0.134+0.841i$ \\
$v_4(0)$ & $-0.751-0.657i$ \\
\hline
\end{tabular}
\end{minipage}
\hfill
\begin{minipage}[t]{0.48\textwidth}
\centering
\captionof{table}{Parameter values in Figure \ref{fig:case1-plot1}.}
\label{tab:case1-params}
\begin{tabular}{c c}
\hline
Parameter & Value \\
\hline
$a_1$ & $0.227$ \\
$a_2$ & $0.0474$ \\
$a_3$ & $0.17$ \\
$a_4$ & $0.874$ \\
$a_{1,1}$ & $-1.17$ \\
$a_{1,2}$ & $1.86$ \\
$a_{1,3}$ & $0.475$ \\
$a_{1,4}$ & $1.09$ \\
$a_{2,2}$ & $-1.84$ \\
$a_{2,3}$ & $-0.0376$ \\
$a_{2,4}$ & $-1.19$ \\
$a_{3,3}$ & $0.455$ \\
$a_{3,4}$ & $-0.341$ \\
$a_{4,4}$ & $-1.84$ \\
\hline
\end{tabular}
\end{minipage}

\vspace{1em}

\noindent
The coupling coefficients satisfy $a_{k,j}=a_{j,k}$.

% ---------- APPENDIX ------------
% --------------------------------

\section{Parameters for Figure \ref{fig:case2-plot1}}
\label{app:case2-params}
The input powers are $(1,8.5,0.4,0.01)$ W and phases are $(0.023,0.65,0.62,1.95)$ radians.
The frequency reference is $2\pi c/(1550\,\mathrm{nm})$, with offsets $(-\Omega,\Omega,-3\Omega,3\Omega)$ and $\Omega=2\pi\times100$ GHz.
The propagation constants are $\beta_j=\tfrac12\beta^{(2)}(\omega_j-\omega_{\rm ref})^2$ with their common mean subtracted, where $\beta^{(2)}=-30\,\mathrm{ps}^2/\mathrm{km}$.
These choices preserve frequency conservation and give nonzero phase mismatch.

\begin{table}[H]\centering
\caption{Initial field amplitudes for Figure~\ref{fig:case2-plot1}; $|A_j|^2$ is in watts.}
\label{tab:case2-init}
\begin{tabular}{cc}\hline Variable & Initial value \\ \hline
$A_1(0)$ & $0.99973551+0.02299797i$ \\
$A_2(0)$ & $2.32096317+1.76440641i$ \\
$A_3(0)$ & $0.51474193+0.36747890i$ \\
$A_4(0)$ & $-0.03701808+0.09289597i$ \\
\hline\end{tabular}\end{table}
\begin{table}[H]\centering
\caption{Parameters for Figure~\ref{fig:case2-plot1}; entries are rounded.}
\label{tab:case2-physical}
\begin{tabular}{cc}\hline Parameter & Value \\ \hline
$c$ & $299792458$ [m/s] \\
$n_2$ & $2.6\times10^{-20}$ [$\mathrm m^2/\mathrm W$] \\
$A_{\rm eff}$ & $10\times10^{-12}$ [$\mathrm m^2$] \\
$\beta^{(2)}$ (GVD) & $-3\times10^{-26}$ [$\mathrm s^2/\mathrm m$] \\
$\omega_1$ & $1.21463076\times10^{15}$ [rad/s] \\
$\omega_2$ & $1.21588739\times10^{15}$ [rad/s] \\
$\omega_3$ & $1.21337412\times10^{15}$ [rad/s] \\
$\omega_4$ & $1.21714403\times10^{15}$ [rad/s] \\
$\beta_1$ & $0.0236870506$ [1/m] \\
$\beta_2$ & $0.0236870506$ [1/m] \\
$\beta_3$ & $-0.0236870506$ [1/m] \\
$\beta_4$ & $-0.0236870506$ [1/m] \\
\hline\end{tabular}\end{table}
The common overlap approximation is $f_{j,k}=f_{1,2,3,4}=1/A_{\rm eff}=1.0\times10^{11}\,\mathrm m^{-2}$ for all indices.
The notebook supplies full-precision inputs.
\begingroup
\emergencystretch=2em
\printbibliography
\endgroup

\end{document}